\documentclass[a4paper,12pt]{article}
\pdfoutput=1
\usepackage[utf8]{inputenc}

\usepackage{test,graphicx,caption}
\usepackage[hyphens]{url} 

\usepackage{hyperref}
\hypersetup{colorlinks,citecolor=blue,urlcolor=magenta}
\usepackage{doi}

\usepackage[style=ext-authoryear-comp,
sorting=nyt,
dashed=false, 
maxcitenames=2, 
maxbibnames=99, 
uniquelist=false,
uniquename=false,
useprefix=true,
giveninits=true, 
natbib, 
date=year 
]{biblatex}

\AtBeginRefsection{\GenRefcontextData{sorting=ynt}}
\AtEveryCite{\localrefcontext[sorting=ynt]}

\DeclareFieldFormat{pages}{#1} 
\renewbibmacro{in:}{\ifentrytype{article}{}{\printtext{\bibstring{in}\intitlepunct}}} 

\DeclareFieldFormat[article,inbook,incollection,inproceedings,patent,thesis,unpublished]{titlecase:title}{\MakeSentenceCase*{#1}} 

\usepackage[inline,shortlabels]{enumitem}
\setlist[enumerate,1]{label=(\roman*)}

\usepackage[margin = 1.25in]{geometry}
\usepackage{setspace}
\newcommand{\cF}{\mathcal{F}}
\newcommand{\cH}{\mathcal{H}}
\newcommand{\cR}{\mathcal{R}}
\newcommand{\cX}{\mathcal{X}}
\renewcommand{\Pr}{\operatorname{P}}

\title{General-Purpose Technologies\\ and Stock Market Bubbles\thanks{This paper is a revised version of our January 2025 paper, \citet*[January 14th]{HiranoKishiToda2025}, titled ``Bursting Bubbles in a Macroeconomic Model''. This research was financially supported by Japan Center for Economic Research; Japan Securities Scholarship Foundation; the Joint Usage/Research Center, Institute of Economic Research, Hitotsubashi University (Grant ID: IERPK2515); JSPS KAKENHI Grant Number JP19K13660; and the Kansai University Fund for Domestic and Overseas Research Fund, 2023.}
}
\author{Tomohiro Hirano\thanks{Department of Economics, Royal Holloway, University of London and The Canon Institute for Global Studies. Email: \href{mailto:tomohih@gmail.com}{tomohih@gmail.com}.} \and Keiichi Kishi\thanks{Faculty of Economics, Kansai University. Email: \href{mailto:kishi@kansai-u.ac.jp}{kishi@kansai-u.ac.jp}.} \and Alexis Akira Toda\thanks{Department of Economics, Emory University and Research Institute for Economics and Business Administration, Kobe University. Email: \href{mailto:alexis.akira.toda@emory.edu}{alexis.akira.toda@emory.edu}.}}

\numberwithin{equation}{section}
\numberwithin{lem}{section}
\numberwithin{prop}{section}

\theoremstyle{plain}

\begin{document}
\maketitle

\begin{abstract}
We develop a macro-finance model linking stock price bubbles to a general-purpose technology (GPT), such as information technology and artificial intelligence. Knowledge spillovers differ across production factors, generating unbalanced growth and causing stock prices to outgrow dividends. Under our conditions, the unique equilibrium contains a bubble on dividend-paying stocks even though agents share common beliefs and rationally anticipate its collapse. The probability that spillovers persist affects the bubble's duration but not its existence. When spillovers equalize as the technology matures, the economy reaches balanced growth and the bubble collapses. Through IPO proceeds, the bubble can increase R\&D employment, while accumulated knowledge remains productive afterward. More broadly, balanced growth is a knife-edge property: the restrictions used to obtain it make stock prices and dividends grow at the same rate, thereby ruling out rational bubbles on dividend-paying assets by construction.

\medskip


\medskip

\end{abstract}

\section{Introduction}

Major technological innovations have often been accompanied by extraordinary booms and busts in asset markets. The Mississippi and South Sea Bubbles around 1720 coincided with innovations in Atlantic trade and insurance \citep*{FrehenGoetzmannRouwenhorst2013}; the Railway Mania of the 1840s accompanied the spread of steam power and the railway network; and even the seemingly quaint Bicycle Mania of the 1890s followed the invention of the pneumatic tire, which subsequently became important to the development of automobiles and aircraft \citep[Chs.~4 and 6]{QuinnTurner2020}. In the twentieth century, the U.S. stock market boom of the 1920s developed alongside electrification and the growth of the automobile, aircraft, motion picture, and radio industries, while the dot-com boom accompanied the commercial expansion of the Internet. More recently, sharp increases in the market values of firms associated with artificial intelligence (AI) have prompted debate over whether this historical pattern may be recurring.\footnote{\href{https://www.forbes.com/sites/johnrau/2025/01/03/is-ai-a-boom-bubble-or-con-heres-what-the-evidence-suggests/}{John Rau, \emph{Forbes}, January 3, 2025}.}

The connection is especially striking for general-purpose technologies (GPTs), which improve over time, spread across sectors, and stimulate complementary innovations.\footnote{\label{fn:GPT}\citet{BresnahanTrajtenberg1995} characterize GPTs by three features:
\begin{enumerate*}
    \item pervasiveness,
    \item improvement, and
    \item innovation spanning.
\end{enumerate*}} \citet{Nicholas2008} describes the 1920s as a period of unprecedented technological advance and growth in intangible capital, yet one that ended with the collapse of Radio Corporation of America, the era's emblematic technology stock. \citet[p.~22]{Scheinkman2014} similarly observes that ``asset price bubbles tend to appear in periods of excitement about innovations,'' pointing to the new technology stocks of the 1920s and to Internet firms during the dot-com boom. Electricity and information technology are canonical examples of GPTs.

This historical pattern motivates a theoretical question. Can an asset price bubble---a situation in which the asset price exceeds the fundamental value defined as the expected present discounted value of dividends---be the unique equilibrium outcome even when agents are rational, share common beliefs, and anticipate that the bubble will eventually collapse completely? This question is challenging for two reasons. First, for a dividend-paying asset, dividends enter the asset-pricing recursion and make the no-bubble condition more restrictive than for a pure bubble---an intrinsically worthless asset that pays no dividends, such as fiat money or cryptocurrency. Second, if the collapse date is known, backward induction rules out a positive bubble at all prior dates. Existing models of stochastic bubbles make the collapse date random, but typically study pure bubbles and require their survival probability to be sufficiently high. In this paper, we develop a macro-finance model in which technological innovation gives rise to a rational stock price bubble and a change in the underlying productivity process, rather than an extrinsic sunspot, triggers its anticipated collapse. The probability that strong technological spillovers continue determines the expected duration of the bubble but not whether it exists.

To isolate the asset-pricing mechanism in the most transparent setting, we first study a simple endowment economy (\S\ref{sec:toy}). This setting allows us to characterize the emergence of bubbles without specifying production or innovation, which we introduce in the full model. The economy is populated by overlapping generations of agents who receive endowments when young and trade a dividend-paying asset (Lucas tree) to save for old age. Aggregate uncertainty arises from a transition between two states. In state $u$, endowments and dividends may grow at different rates (\emph{unbalanced growth}). At each date, the economy may switch to state $b$, in which endowments and dividends grow at the same rate (\emph{balanced growth}). State $b$ is absorbing: once the economy reaches balanced growth, it remains there indefinitely. The equilibrium asset price is unique, and the asset price equals its fundamental value in state $b$. Theorem \ref{thm:bubble_toy} gives necessary and sufficient conditions for the existence of a bubble in state $u$. When relative risk aversion is below one, these conditions have a simple interpretation: while state $u$ persists, endowments outgrow dividends, so the price-dividend ratio diverges; upon transition to state $b$, consumption falls sufficiently relative to the level that would obtain if state $u$ continued. Lower consumption after the transition raises the marginal value of consumption, but this increase is weaker than the decline in the stock payoff. If and only if both conditions hold, a bubble that is expected to collapse emerges as the unique equilibrium outcome. Finite lifetimes are not essential: \S\ref{sec:infinitely-lived} establishes an analogous result in a model with infinitely lived agents and continual cohort entry.

We next apply this result in a macro-finance model of innovation and intangible capital (\S\ref{sec:innovation}). The production structure is a variety-expansion model in the tradition of \citet[Ch.~3]{GrossmanHelpman1991book}. Skilled agents choose between producing knowledge-intensive intermediate goods and conducting R\&D, whereas unskilled agents work in the final-good sector. Agents who conduct R\&D create new varieties and establish firms whose shares are claims to monopoly profits. Knowledge raises the productivities of both knowledge-intensive inputs and unskilled labor. The stock of product varieties, $n_t$, measures accumulated knowledge, and the productivity of R\&D is proportional to this stock. These specifications represent, respectively, the pervasiveness, improvement, and innovation-spanning features of a GPT.

Suppose that final output is produced using a constant elasticity of substitution (CES) technology, with an elasticity of substitution between the knowledge-intensive input and unskilled labor below one, consistent with empirical evidence \citep{OberfieldRaval2021,GechertHavranekIrsovaKolcunova2022}. In state $u$, knowledge has a stronger spillover effect on the productivity of the knowledge-intensive input than on unskilled-labor productivity. If relative risk aversion is below one, Theorem \ref{thm:bubble_GH} shows that the unique equilibrium stock price contains a bubble throughout state $u$, even though all agents anticipate its eventual collapse. Conditional on remaining in state $u$, the price-dividend ratio asymptotically grows exponentially. Unlike models in which bubbles are attached to lower-growth assets such as land \citep{HiranoToda2025PNAS}, the bubble here is attached to a higher-growth asset: equity issued by intermediate-good firms. Upon transition to state $b$, the bubble collapses and the price-dividend ratio becomes constant. Because the transition does not reduce the stock of accumulated knowledge, if knowledge remains productive after the transition, a later transition raises output and both wages at every fixed post-transition horizon.
Together with the R\&D response to bubble-backed IPO proceeds studied in \S\ref{sec:macro_consequences}, this result identifies a potential productive legacy of a technology-driven bubble: the bubble can stimulate innovation before it collapses, while the resulting knowledge remains productive afterward.\footnote{\label{fn:Bezos}Jeff Bezos makes a similar distinction between industrial and financial bubbles: ``The ones that are industrial are not nearly as bad. They could even be good because when the dust settles and you see who are the winners, society benefits from those inventions.'' Bezos, in conversation with John Elkann at Italian Tech Week, Turin, October 3, 2025. \href{https://www.youtube.com/watch?v=HB3Rq6ZLliM&t=2260s}{Video recording}.}

A broader implication concerns the emphasis of macroeconomics on balanced growth. Balanced growth is itself a knife-edge property, requiring restrictions on technology and the direction of technical change. Proposition \ref{prop:knife-edge} makes the asset-pricing consequence explicit: in our model, the required restriction takes one of two forms, Cobb--Douglas production or equal knowledge spillovers across factors. Either makes stock prices and dividends grow at the same rate. Because they make the price-dividend ratio constant, these restrictions eliminate the relative-growth conditions that generate the bubble. Thus, constructing a model to exhibit balanced growth (by implicitly introducing knife-edge restrictions) rules out rational bubbles attached to dividend-paying assets \emph{by construction}. Relaxing the knife-edge restrictions, even temporarily, produces unbalanced growth and fundamentally changes the asset-pricing implications. This result is an asset-pricing analogue of the Uzawa steady-state growth theorem.

\paragraph{Related literature}

Our paper contributes to the rational bubble literature pioneered by \citet{Samuelson1958}, \citet{Bewley1980}, \citet{Tirole1985}, \citet{ScheinkmanWeiss1986}, \citet{Kocherlakota1992}, and \citet{SantosWoodford1997}; see \citet{HiranoToda2024JME} and \citet{Barlevy2025} for recent reviews. Most of this literature studies pure bubbles, partly because the Bubble Impossibility Theorem of \citet{SantosWoodford1997} rules out bubbles when dividends are non-negligible relative to aggregate endowment. Despite their usefulness, pure bubble models have three limitations for applications to stocks, land, and housing \citep[\S4.7]{HiranoToda2024JME}. First, pure bubble assets pay no dividends, so these models do not directly describe bubbles attached to dividend-paying assets. Second, they often admit a continuum of equilibria, so quantitative and policy conclusions may depend on equilibrium selection. Third, the price-dividend ratio used in the econometric bubble-detection literature is undefined, preventing a direct connection between the theory and empirical measures of bubbles \citep{PhillipsShiYu2015,PhillipsShi2018,PhillipsShi2020}.

Recent work develops rational bubbles attached to real, dividend-paying assets. \citet{HiranoToda2025JPE} establish a Bubble Necessity Theorem under which every equilibrium contains a bubble of non-negligible size. Although \citet{Tirole1985} claimed that bubbles may be necessary under a low interest rate condition, \citet{PhamTodaComment} show that this claim is incorrect: bubble necessity also depends on initial conditions. \citet{HiranoToda2025PNAS} generate recurrent expansions and contractions in land bubbles under aggregate uncertainty. In the two Hirano--Toda models, a bubble, once present, never collapses. \citet{Sorger2026} provides a recent example involving dividend-paying stocks: he derives conditions under which permanent stock price bubbles are inevitable in variants of the Diamond and Romer-style AK models. Like these papers, we establish conditions for bubble necessity; unlike them, we study a bubble whose complete collapse is anticipated. The bubble is attached to stock issued by innovative firms, exists during unbalanced growth, and disappears upon transition to an absorbing balanced-growth state.

The literature on collapsing rational bubbles originates with \citet{Blanchard1979}. In the general-equilibrium model of \citet{Weil1987}, a stochastic pure bubble can exist only if its survival probability is sufficiently high; \citet{MichelWigniolle2003} construct temporary monetary bubbles; and \citet{Guerron-QuintanaHiranoJinnai2023} generate recurrent bubbles and crashes through changes in investor sentiment in a growth model. Our model differs in three respects: the asset pays dividends, the bubble collapses when the underlying productivity process changes rather than in response to an extrinsic sunspot, and the survival probability affects expected duration but not the existence conditions.

Our paper is also related to technological innovation and asset booms. \citet{Olivier2000} shows that rational equity bubbles can increase firm creation and growth. However, \citet{Olivier2000} permits assets with identical dividend streams to trade at different prices, whereas we impose the law of one price as in ordinary macro-finance models. \citet{Zeira1999} and \citet{PastorVeronesi2009} generate stock-price increases and collapses through Bayesian learning about dividend growth or technology profitability; prices therefore remain equal to fundamental values. \citet{HaddadHoLoualiche2022} document that innovation becomes overvalued during booms and explain the pattern through investor disagreement. In our model, agents have common beliefs, the stock price exceeds the expected present discounted value of dividends, and the bubble component of IPO proceeds increases R\&D employment.

Finally, our interpretation of a GPT builds on the pervasiveness, improvement, and innovation-spanning framework of \citet{BresnahanTrajtenberg1995}. The gradual diffusion of GPTs through complementary investment and organizational change is documented by \citet{JovanovicRousseau2005} and \citet{BrynjolfssonRockSyverson2021}, while \citet{CockburnHendersonStern2019} emphasize AI as a general-purpose method of invention. In the model, the dependence of both factor productivities on knowledge represents pervasiveness, knowledge accumulation represents improvement, and the proportionality of R\&D productivity to the knowledge stock represents innovation spanning. The result on balanced growth relates to the Uzawa theorem and its modern extensions \citep{Uzawa1961,Schlicht2006,JonesScrimgeour2008,GrossmanHelpmanOberfieldSampson2017,CaseyHorii2024}: in our model, the restrictions that imply balanced growth also make the price-dividend ratio constant and eliminate the relative-growth conditions used in our bubble theorem.

\section{Bubble necessity despite anticipated collapse}\label{sec:toy}

To isolate the asset-pricing conditions for stochastic bubbles, we present a simple endowment economy and identify conditions under which a bubble expected to collapse is the unique equilibrium outcome.

\subsection{Model}\label{subsec:toy_model}

We consider a variant of the \citet{Lucas1978} economy. Time is indexed by $t=0,1,\dotsc$. Uncertainty is resolved according to a filtration $\set{\cF_t}_{t=0}^\infty$ on a probability space $(\Omega,\cF,\Pr)$. There is a single consumption good at each date, which serves as the num\'eraire. Markets are complete: at each date, agents can trade a complete set of one-period Arrow securities in zero net supply.

\paragraph{Agents and preferences}

We use a standard two-period overlapping generations (OLG) structure. At $t=0$, there is a unit measure of initial old, who consume and exit. At every date $t\ge0$, a unit measure of young agents is born and lives for two periods. This is the simplest setting in which successive generations trade a long-lived asset.

For analytical tractability, we suppose that agents have Epstein-Zin preferences with elasticity of intertemporal substitution (EIS) equal to 1.\footnote{More precisely, \eqref{eq:utility} is the two-period specialization of homogeneous Kreps-Porteus/Epstein-Zin recursive utility with unit EIS. To see this, let $V_t$ denote utility in levels. The unit-EIS recursion can be written as $V_t=(c_t^y)^{1-\beta}\E_t[V_{t+1}^{1-\gamma}]^\frac{\beta}{1-\gamma}$. Imposing the terminal condition $V_{t+1}=c_{t+1}^o$ and defining $U_t\coloneqq\log V_t$ yields \eqref{eq:utility}. \citet[p.~1212]{Blanchard2019} uses the identical specification in a two-period OLG model.} We treat the case with general EIS in Appendix \ref{app:non-unit_EIS}.

\begin{asmp}\label{asmp:utility}
The utility function of an agent born at date $t$ is given by
\begin{equation}
    U(c_t^y,c_{t+1}^o)=(1-\beta)\log c_t^y+\beta\begin{cases*}
        \frac{1}{1-\gamma}\log \E_t[(c_{t+1}^o)^{1-\gamma}] & if $\gamma\neq 1$,\\
        \E_t[\log c_{t+1}^o] & if $\gamma=1$,
    \end{cases*}
    \label{eq:utility}
\end{equation}
where $\beta\in (0,1)$ is the discount factor, $\gamma\ge 0$ is the relative risk aversion, $(c_t^y,c_{t+1}^o)$ denote consumption when young and old, and $\E_t[\cdot]\coloneqq \E[\cdot|\cF_t]$ denotes the expectation conditional on information at date $t$.
\end{asmp}

\paragraph{Endowments and dividends}

At date $t\ge 0$, the young are endowed with $e_t>0$ units of the good, whereas the old are not.\footnote{The assumption of zero non-financial income for the old is imposed only for analytical tractability and is not essential. This assumption is adopted by \citet{Tirole1985}, \citet*{AllenBarlevyGale2025}, and \citet{HiranoToda2025PNAS}, among others.} The initial old are endowed with a unit supply of a long-lived asset (tree), which produces $D_t>0$ units of the good as dividend at date $t$.\footnote{We can think of the asset as the market portfolio. Introducing multiple assets will not change the results because only the young purchase the assets and they are homogeneous, so they make the same portfolio choice, ending up demanding the market portfolio. See $\S2.6$ of our first version of the present paper, \citet*{HiranoKishiToda2025}, which analyzes the case with multiple saving vehicles.} The sequence $\set{(e_t,D_t)}_{t=0}^\infty$ follows some stochastic process adapted to $\set{\cF_t}_{t=0}^\infty$. At this point, we do not need any restriction on this stochastic process.

\paragraph{Budget constraints}

Because only identical young agents trade one-period Arrow securities, symmetry and zero net supply imply zero equilibrium holdings. Their prices are nevertheless uniquely determined by the young agents' intertemporal marginal rate of substitution, represented by the SDF $m_{t\to t+1}$ derived in \eqref{eq:SDF}. We therefore suppress Arrow security prices and positions from the notation and write the equilibrium budget constraints in reduced form:
\begin{subequations}\label{eq:budget}
\begin{align}
    &\text{Young:} & c_t^y+Q_t\theta_t&=e_t, \label{eq:budget_young}\\
    &\text{Old:} & c_{t+1}^o&=(Q_{t+1}+D_{t+1})\theta_t, \label{eq:budget_old}
\end{align}
\end{subequations}
where $Q_t>0$ is the ex-dividend asset price at date $t$ and $\theta_t$ is the asset holdings of generation $t$.

\paragraph{Equilibrium}

\begin{defn}\label{defn:eq_toy}
The stochastic process $\set{(Q_t,c_t^y,c_t^o,\theta_t)}_{t=0}^\infty$ is a \emph{rational expectations equilibrium} if
\begin{enumerate}
    \item (Utility maximization) the initial old consume $c_0^o=Q_0+D_0$; for each $t\ge 0$, $(c_t^y,\theta_t,c_{t+1}^o)$ maximizes the utility \eqref{eq:utility} subject to the budget constraints \eqref{eq:budget},
    \item (Commodity market clearing) for each $t$, we have $c_t^y+c_t^o=e_t+D_t$,
    \item (Asset market clearing) for each $t$, we have $\theta_t=1$.
\end{enumerate}
\end{defn}

As is well known, due to unit EIS, the optimal consumption of the young is $c_t^y=(1-\beta)e_t$.\footnote{\label{fn:log_consumption}To see this, letting $c=c_t^y$, $e=e_t$, and $R=(Q_{t+1}+D_{t+1})/Q_t$ be the gross return on the asset, the optimization problem reduces to maximizing $(1-\beta)\log c+\frac{\beta}{1-\gamma}\log \left(\E[(R(e-c))^{1-\gamma}]\right)=(1-\beta)\log c+\beta \log (e-c)+\text{constant}$. Maximizing over $c$ yields $c=(1-\beta)e$.} Then, using the young's budget constraint \eqref{eq:budget_young} and the asset market clearing condition $\theta_t=1$, we obtain the asset price $Q_t=\beta e_t$. Therefore, we obtain the following equilibrium existence and uniqueness result.

\begin{prop}\label{prop:eq_toy}
There exists a unique rational expectations equilibrium. The asset price is $Q_t=\beta e_t$ and consumption is $(c_t^y,c_t^o)=((1-\beta)e_t,\beta e_t+D_t)$.
\end{prop}

\subsection{Asset pricing and bubbles}\label{subsec:toy_SDF}\label{subsec:toy_bubble}

To characterize asset pricing in this equilibrium, first note that the one-period stochastic discount factor (SDF) implied by \eqref{eq:utility} is
\begin{equation}
    m_{t\to t+1}=\frac{\partial U/\partial c_{t+1}^o}{\partial U/\partial c_t^y}=\frac{\beta}{1-\beta}\frac{c_t^y(c_{t+1}^o)^{-\gamma}}{\E_t[(c_{t+1}^o)^{1-\gamma}]}. \label{eq:SDF}
\end{equation}
The expression is also valid for $\gamma=1$. Substituting the equilibrium quantities in Proposition \ref{prop:eq_toy} gives
\begin{equation}
    m_{t\to t+1}=\frac{\beta}{1-\beta}\frac{(1-\beta)e_t (\beta e_{t+1}+D_{t+1})^{-\gamma}}{\E_t[(\beta e_{t+1}+D_{t+1})^{1-\gamma}]}=\frac{Q_t(Q_{t+1}+D_{t+1})^{-\gamma}}{\E_t[(Q_{t+1}+D_{t+1})^{1-\gamma}]}. \label{eq:SDF_Q}
\end{equation}
The last expression depends only on risk aversion and the conditional distribution of the asset payoff.

We now use this SDF to decompose the asset price into its fundamental and bubble components. Following the standard rational bubble literature, the fundamental value is the expected present discounted value of dividends, and the bubble is the difference between the asset price and this value.\footnote{See \citet{SantosWoodford1997} and \citet{HiranoToda2024JME,HiranoToda2025JPE,HiranoToda2025EJW} for more discussion.} Because young agents hold the asset in equilibrium, their Euler equation implies
\begin{equation}
    Q_t=\E_t[m_{t\to t+1}(Q_{t+1}+D_{t+1})]. \label{eq:noarbitrage}
\end{equation}
For $T>t$, let $m_{t\to T}\coloneqq \prod_{s=t}^{T-1}m_{s\to s+1}$ be the stochastic discount factor between dates $t$ and $T$. Iterating \eqref{eq:noarbitrage} and applying the law of iterated expectations yields
\begin{equation}
    Q_t=\E_t\left[\sum_{s=t+1}^T m_{t\to s}D_s\right]+\E_t[m_{t\to T}Q_T]. \label{eq:Q_iter}
\end{equation}
Because $m_{t\to s}>0$ and $D_s\ge0$, the first term on the right-hand side of \eqref{eq:Q_iter} is increasing in $T$ and bounded above by $Q_t$. Applying the monotone convergence theorem as $T\to\infty$ gives the decomposition
\begin{equation}
    Q_t=\underbrace{\E_t\left[\sum_{s=t+1}^\infty m_{t\to s}D_s\right]}_\text{fundamental value $V_t$}+\underbrace{\lim_{T\to\infty}\E_t[m_{t\to T}Q_T]}_\text{bubble $B_t$}. \label{eq:decompose}
\end{equation}
The bubble is nonnegative because the second term in \eqref{eq:Q_iter} is nonnegative. Hence, the asset price equals its fundamental value if and only if the no-bubble condition
\begin{equation}
    \lim_{T\to\infty}\E_t[m_{t\to T}Q_T]=0 \label{eq:nobubble}
\end{equation}
holds.\footnote{Some authors such as \citet{SantosWoodford1997}, \citet{Montrucchio2004}, and \citet{HiranoToda2025JPE} use the term ``transversality condition'' instead of ``no-bubble condition''. We use the latter to avoid confusion between the transversality condition \emph{for asset pricing} and for \emph{optimality} (in infinite-horizon optimal control problems), which are completely different.} Economically, $B_t$ is the part of the price sustained by expected future resale rather than dividends. When $B_t=0$, the expected dividend stream fully supports the price; when $B_t>0$, the price contains an additional speculative component.\footnote{\citet[p.~1]{Kaldor1939} defines speculation as ``the purchase (or sale) of goods with a view to re-sale (re-purchase) at a later date''.}

\subsection{Temporary unbalanced growth and stochastic bubbles}

Recent work by some of us links bubbles to \emph{unbalanced growth}, in which sectors or factors grow at different rates \citep{HiranoToda2024JME,HiranoToda2025JPE,HiranoToda2025PNAS}. With the preceding asset-pricing decomposition in hand, we ask whether bubbles remain possible, or even necessary, when unbalanced growth is temporary. We introduce a regime-switching process with an absorbing return to balanced growth.

\begin{asmp}\label{asmp:states}
There are two states of the economy denoted by $u,b$. Letting $z_t\in \set{u,b}$ denote the state at date $t$, the transition probabilities are given by
\begin{align*}
    \Pr[z_{t+1}=u \mid z_t=u] &=\pi \in (0,1), \\
    \Pr[z_{t+1}=b \mid z_t=b] &=1.
\end{align*}
\end{asmp}

Assumption \ref{asmp:states} implies that state $b$ is absorbing.

\begin{asmp}\label{asmp:balanced}
If state $b$ is reached at date $\tau$, all subsequent endowments and dividends are deterministic, and they grow at the same rate: $e_{t+1}/e_t=D_{t+1}/D_t$ for all $t\ge \tau$.
\end{asmp}

Assumption \ref{asmp:balanced} implies that once state $b$ is reached, all uncertainty is resolved and endowments and dividends subsequently share a common, possibly history-dependent gross growth factor (balanced growth). The following proposition shows that the asset price then equals its fundamental value.

\begin{prop}\label{prop:nobubble_b}
If Assumptions \ref{asmp:utility}--\ref{asmp:balanced} hold and state $b$ is reached at date $\tau$, then $Q_t=V_t$ for all $t\ge\tau$.
\end{prop}

To see the intuition, suppose $z_0=b$ and endowments and dividends have the common gross growth factor $G>0$. Because $Q_t=\beta e_t$, the asset price has the same gross growth factor $G$, whereas its gross return is
\begin{equation*}
    R\coloneqq \frac{Q_{t+1}+D_{t+1}}{Q_t}=G\frac{\beta e_0+D_0}{\beta e_0}>G.
\end{equation*}
Thus, the discount rate exceeds price growth, the discounted terminal price vanishes, and the asset is bubbleless.

By Proposition \ref{prop:nobubble_b}, bubbles cannot arise in state $b$. Therefore, in what follows, we assume $z_0=u$. To check the no-bubble condition \eqref{eq:nobubble}, we evaluate the expectation $\E_0[m_{0\to t}Q_t]$. To simplify the calculation, let $\tau$ be the stopping time that represents the first date at which the economy switches to state $b$: $\tau=\min\set{t:z_t=b}$. By Assumption \ref{asmp:states}, $\tau$ is geometrically distributed with probability
\begin{equation}
    \Pr[\tau=j]=\pi^{j-1}(1-\pi) \label{eq:p_geometric}
\end{equation}
for $j=1,2,\dotsc$. Then we obtain
\begin{equation}
    \E_0[m_{0\to t}Q_t]=\sum_{j=1}^t\pi^{j-1}(1-\pi)\E_0[m_{0\to t}Q_t \mid \tau=j]+\pi^t\E_0[m_{0\to t}Q_t \mid \tau>t]. \label{eq:mQ_iter}
\end{equation}
The following proposition provides a necessary and sufficient condition for the nonexistence of bubbles.

\begin{prop}\label{prop:nobubble_toy}
Suppose Assumptions \ref{asmp:utility}--\ref{asmp:balanced} hold. Then there is no bubble if and only if
\begin{equation}
    \lim_{t\to\infty}\pi^t\E_0[m_{0\to t}Q_t \mid \tau>t]=0. \label{eq:TVC}
\end{equation}
If $\gamma=1$ or $(e_t,D_t)$ is known at date $t-1$ for all $t\ge 1$, then there is no bubble.  
\end{prop}

The intuition for Proposition \ref{prop:nobubble_toy} is the following. Because state $b$ is bubbleless by Proposition \ref{prop:nobubble_b}, paths on which the economy has already switched to $b$ make an asymptotically negligible contribution to the discounted terminal price. Thus, only paths that remain in state $u$ can sustain a bubble, yielding condition \eqref{eq:TVC}.

Proposition \ref{prop:nobubble_toy} also shows that new information about endowments or dividends and $\gamma\neq1$ are necessary for a bubble that eventually collapses completely. In particular, the commonly used logarithmic utility ($\gamma=1$) does not work, which is precisely why we imposed the Epstein-Zin specification in Assumption \ref{asmp:utility}. This contrasts with the model of \citet{HiranoToda2025PNAS} that features aggregate uncertainty, where bubbles are possible under log utility because the economy never enters an absorbing balanced growth path. To characterize the remaining case, we impose a simple conditional structure on endowments and dividends.

\begin{asmp}\label{asmp:conditional}
For every $t\ge1$, the past regime history $(z_0,\dots,z_{t-1})$ determines two possible endowment-dividend pairs, $(e_t^u,D_t^u)$ and $(e_t^b,D_t^b)$. If the current state is $z_t=z\in\set{u,b}$, then $(e_t,D_t)=(e_t^z,D_t^z)$. There are no other sources of aggregate uncertainty.
\end{asmp}

Assumption \ref{asmp:conditional} allows endowments and dividends to respond to the current regime. Given the past regime history, the realization of $z_t$ is the only new aggregate information arriving at date $t$. Throughout, for any variable $x_t$ and state $z\in\set{u,b}$, we let $x_t^z$ denote its value conditional on $z_0=\dots=z_{t-1}=u$ and $z_t=z$. We can now characterize the existence of stochastic bubbles.

\begin{thm}[Bubble characterization with anticipated collapse]\label{thm:bubble_toy}
Suppose Assumptions \ref{asmp:utility}--\ref{asmp:conditional} hold. For $z\in\set{u,b}$, let $\Pi_t^z\coloneqq Q_t^z+D_t^z=\beta e_t^z+D_t^z$ denote the stock payoff and let $Q_t^z\coloneqq \beta e_t^z=\Pi_t^z-D_t^z>0$. If $z_0=u$, then there is a bubble at $t=0$ if and only if
\begin{equation}
    \sum_{t=1}^\infty \frac{D_t^u}{Q_t^u}<\infty
    \quad\text{and}\quad
    \sum_{t=1}^\infty \left(\frac{\Pi_t^b}{\Pi_t^u}\right)^{1-\gamma}<\infty. \label{eq:finite_sum}
\end{equation}
\end{thm}

We prove Theorem \ref{thm:bubble_toy} in the main text, as its proof is helpful for understanding why the assumptions are essential.

\begin{proof}
Let $\widehat Q_t\coloneqq m_{0\to t}Q_t$ and define the discounted value associated with remaining in state $u$ through date $t$ by $S_t\coloneqq \pi^t\E_0[\widehat Q_t\mid \tau>t]$. Using the SDF \eqref{eq:SDF_Q}, we obtain
\begin{equation*}
    \widehat Q_t=Q_0\prod_{s=1}^t \frac{Q_s(Q_s+D_s)^{-\gamma}}{\E_{s-1}[(Q_s+D_s)^{1-\gamma}]}.
\end{equation*}
Conditional on $\tau>t$, Assumptions \ref{asmp:states} and \ref{asmp:conditional} imply that
\begin{align}
    S_t
    &=Q_0\prod_{s=1}^t \frac{\pi Q_s^u(\Pi_s^u)^{-\gamma}}{\pi (\Pi_s^u)^{1-\gamma}+(1-\pi) (\Pi_s^b)^{1-\gamma}} \notag \\
    &=Q_0\prod_{s=1}^t \frac{1}{1+D_s^u/Q_s^u + \frac{1-\pi}{\pi}(\Pi_s^u/Q_s^u)(\Pi_s^b/\Pi_s^u)^{1-\gamma}}. \label{eq:pimQ}
\end{align}
Define
\begin{equation}
    a_s\coloneqq D_s^u/Q_s^u+\frac{1-\pi}{\pi}(\Pi_s^u/Q_s^u)(\Pi_s^b/\Pi_s^u)^{1-\gamma}. \label{eq:as}
\end{equation}
Because $a_s\ge 0$, \eqref{eq:pimQ} and the inequalities $1+a\le \exp(a)$ and $\prod_{s=1}^t(1+a_s)\ge 1+\sum_{s=1}^t a_s$ imply
\begin{equation}
    Q_0\exp\left(-\sum_{s=1}^t a_s\right)
    \le S_t
    \le Q_0\left(1+\sum_{s=1}^t a_s\right)^{-1}. \label{eq:pimQ_bounds}
\end{equation}
Letting $t\to\infty$ in \eqref{eq:pimQ_bounds}, we have $\lim_{t\to\infty}S_t>0$ if and only if $\sum_{s=1}^\infty a_s<\infty$. By Proposition \ref{prop:nobubble_toy}, there is a bubble at $t=0$ if and only if $\sum_{s=1}^\infty a_s<\infty$.

It remains to express this condition in terms of the two conditions in \eqref{eq:finite_sum}. If $\sum_{s=1}^\infty a_s<\infty$, then the first condition follows from $0\le D_s^u/Q_s^u\le a_s$. Moreover, since $\Pi_s^u/Q_s^u=1+D_s^u/Q_s^u\ge 1$, \eqref{eq:as} implies
\begin{equation*}
    (\Pi_s^b/\Pi_s^u)^{1-\gamma}\le \frac{\pi}{1-\pi}a_s,
\end{equation*}
so the second condition in \eqref{eq:finite_sum} holds.

Conversely, suppose the two conditions in \eqref{eq:finite_sum} hold. Then $D_s^u/Q_s^u\to 0$, so the sequence $\Pi_s^u/Q_s^u=1+D_s^u/Q_s^u$ is bounded. Therefore, both terms on the right-hand side of \eqref{eq:as} are summable, and hence $\sum_{s=1}^\infty a_s<\infty$. \qedhere
\end{proof}

Theorem \ref{thm:bubble_toy} is a stochastic counterpart of the ``Bubble Characterization Lemma'' \citep[Proposition 7]{Montrucchio2004}. Without aggregate uncertainty, Montrucchio's condition coincides with the first condition in \eqref{eq:finite_sum}. Under our regime-switching structure, the first condition governs the bubble while state $u$ persists, whereas the second accounts for its complete collapse upon transition to state $b$.

This simple model yields three economic implications.

\begin{enumerate}
    \item\label{item:imp1}
    While state $u$ persists, the first condition in \eqref{eq:finite_sum} drives the dividend yield to zero and the price-dividend ratio to infinity, so asset returns increasingly consist of capital gains. The bubble is therefore a temporary departure from the balanced growth path, where the asset price equals its fundamental value.

    \item\label{item:imp2}
    Because the equilibrium asset price is unique, whenever the conditions in Theorem \ref{thm:bubble_toy} hold, a bubble expected to collapse is the unique equilibrium outcome. In this sense, its emergence is a necessity, not merely a possibility.

    \item\label{item:imp3}
    When $\gamma<1$, the two conditions in \eqref{eq:finite_sum} have a simple economic interpretation:
    \begin{enumerate}
        \item\label{item:imp3a}
        during unbalanced growth, endowments outgrow dividends; and
        \item\label{item:imp3b}
        upon transition to balanced growth, old-age consumption falls sharply enough relative to its continuation-state level.\footnote{In this economy, old-age consumption equals the stock payoff. The transition raises the SDF by lowering old-age consumption. When $\gamma<1$, however, this response is weaker than the fall in the stock payoff, so the transition-state contribution $(\Pi_t^b/\Pi_t^u)^{1-\gamma}$ becomes small.}
    \end{enumerate}
    Thus, the bubble can survive while the high-growth state persists only if its eventual collapse becomes sufficiently severe.
\end{enumerate}

\begin{rem}[Role of risk aversion]
In the multiperiod Epstein-Zin specification, let $\psi>0$ denote the EIS. Then $\gamma<1/\psi$ implies a preference for late resolution of uncertainty, so $\gamma<1$ has this interpretation under unit EIS. Appendix \ref{app:non-unit_EIS} extends the characterization to non-unit EIS and shows that the condition remains $\gamma<1$, so the restriction does not arise from unit EIS. The condition $\gamma<1$ is substantive for obtaining complete collapse. When $\Pi_t^b/\Pi_t^u\to0$, as in our innovation model, the second condition in \eqref{eq:finite_sum} can hold only if $\gamma<1$: at $\gamma=1$, every summand equals one, whereas for $\gamma>1$, the summands diverge. Economically, the decline in the transition-state payoff must dominate the increase in its marginal valuation. 
\end{rem}

\begin{rem}[Comparison with Weil's stochastic bubble]\label{rem:Weil}
Theorem \ref{thm:bubble_toy} differs fundamentally from the stochastic pure bubble of \citet{Weil1987}. In Weil's stationary OLG economy, a positive stationary bubble that survives each period with probability $\pi$ exists only if $\pi\beta a>(1-\beta)b$, where $a$ and $b$ are the endowments of the young and old \citep[\S4.1]{HiranoToda2024JME}. A sufficiently high collapse probability therefore rules it out. By contrast, the conditions in \eqref{eq:finite_sum} are independent of $\pi$, which enters the proof only through the positive constant $(1-\pi)/\pi$ and hence cannot affect summability. Weil's bubble is a pure bubble whose collapse is unrelated to fundamentals (it is a sunspot); here, the dividend-paying asset remains viable in both regimes, while the transition changes endowment and dividend dynamics and eliminates the bubble component. Thus, $\pi$ governs the bubble's expected duration, not its existence.
\end{rem}

Unlike the permanent bubbles in \citet{HiranoToda2025JPE,HiranoToda2025PNAS}, the bubble here collapses at an uncertain transition date; a known terminal date would rule it out by backward induction. Permanent bubbles require only condition \ref{item:imp3a}, whereas complete collapse also requires condition \ref{item:imp3b}: old-age consumption must fall sufficiently at the transition. The next section develops a production economy that generates both conditions.

\section{Model of innovation and intangible capital} \label{sec:innovation}

\S\ref{sec:toy} established bubble necessity despite anticipated collapse in an endowment economy. We now embed those conditions in a production economy with innovation and intangible capital. This section develops and solves the model; \S\ref{sec:GH_bubble} gives conditions under which a stochastic stock price bubble for intermediate-good firms is the unique equilibrium outcome. As in \S\ref{sec:toy}, one-period Arrow securities are in zero net supply and have zero equilibrium holdings. From this point onward, we suppress their prices and positions and focus on the stock market.

\subsection{Model}\label{subsec:innovation_model}

The essential structure of the model is similar to the variety expansion model of \citet[Ch.~3]{GrossmanHelpman1991book}, which builds on the monopolistic competition model of \citet{DixitStiglitz1977}. To study the emergence of stochastic stock price bubbles, we reformulate their model as an overlapping generations economy in the spirit of \S\ref{sec:toy}.\footnote{In \S\ref{sec:infinitely-lived}, we replace the two-period OLG structure with infinitely lived agents and show that the bubble result continues to hold. Although we employ a widely adopted monopolistic competition model, a stock price bubble could also be modeled under perfect competition.}

\paragraph{Agents}

At each date, continua of skilled (high-skilled or human-capital type) and unskilled (low-skilled or labor type) agents of masses $H>0$ and $L>0$ are born. Agents live for two periods and are endowed with one unit of their respective type of labor when young. As in \S\ref{sec:toy}, agents have rational expectations and lifetime utility \eqref{eq:utility}.

\paragraph{Timing}

At the beginning of date $t$, factor productivities are realized, and $n_t$ intermediate-good varieties are available for production. Young agents supply labor; skilled agents choose between producing intermediate goods and conducting R\&D, whereas unskilled agents work in the final-good sector. Production and R\&D then take place. The $n_t$ incumbent firms pay dividends to their old shareholders, and R\&D creates $n_{t+1}-n_t$ new firms, which begin producing at date $t+1$. After current dividends have been paid, the old sell their ex-dividend shares in incumbent firms, while innovators issue shares in the newly created firms through IPOs. The young purchase all $n_{t+1}$ shares and carry them into date $t+1$.

\paragraph{Final good sector}

A representative competitive firm produces the final good, which is used for consumption and serves as the num\'eraire, using the aggregate production function
\begin{equation}
    Y_t = F(A_{Xt}X_t,A_{Lt}L_t),    \label{eq:prod_Y}
\end{equation}
where $Y_t$ is the output of the final good, $X_t$ and $L_t$ are the inputs of the knowledge-intensive good and unskilled labor, and $A_{Xt}$ and $A_{Lt}$ are their factor-augmenting productivities. We assume that the function $F(X,L)$ is neoclassical.

\begin{asmp}\label{asmp:F}
$F:\R_{++}^2\to \R_{++}$ is continuously differentiable, concave, homogeneous of degree one, and has strictly positive partial derivatives. Furthermore, letting $g(x)\coloneqq (F_X/F_L)(x,1)$ be the marginal rate of technical substitution (where $F_X,F_L$ denote the partial derivatives with respect to $X,L$), we have $g((0,\infty))=(0,\infty)$.
\end{asmp}

The last condition on $g$ is an Inada-type condition. A typical example satisfying Assumption \ref{asmp:F} is the constant elasticity of substitution (CES) specification
\begin{equation}
    F(X,L)=\begin{cases*}
        \left(\alpha X^{1-\rho} + (1-\alpha) L^{1-\rho} \right)^{\frac{1}{1-\rho}} & if $0<\rho\neq 1$,\\
        X^\alpha L^{1-\alpha} & if $\rho=1$,
    \end{cases*} \label{eq:F_CES}
\end{equation}
where $\rho>0$ is the reciprocal of the elasticity of substitution between the two inputs and $\alpha\in (0,1)$.\footnote{The setting in \citet[Ch.~3]{GrossmanHelpman1991book} is obtained by setting $F(X,L)=X$ and interpreting \eqref{eq:prod_Y} as a utility function rather than a production function: see \citet[Footnote 2]{Benassy1998}. When interpreted as a utility function, agents derive utility from two types of goods, $A_{Xt}X_t$ and $A_{Lt}L$.} Then straightforward algebra shows $F_X=\alpha F^\rho X^{-\rho}$ and $F_L=(1-\alpha) F^\rho L^{-\rho}$, so
\begin{equation}
    g(x)=(F_X/F_L)(x,1)=\frac{\alpha}{1-\alpha}x^{-\rho}, \label{eq:g_CES}
\end{equation}
which maps $(0,\infty)$ onto $(0,\infty)$.

Taking as given the price of the knowledge-intensive good $P_t>0$ and the wage of unskilled labor $w_{Lt}>0$, the final-good firm maximizes the profit
\begin{equation}
    Y_t-P_tX_t-w_{Lt}L_t. \label{eq:profit_Y}
\end{equation}
Because the production function \eqref{eq:prod_Y} is homogeneous of degree one, the maximized profit is zero.

\paragraph{Knowledge-intensive good sector}

A representative competitive firm produces the knowledge-intensive good using the aggregate production function
\begin{equation}
    X_t = n_t^{-\frac{1}{\sigma-1}}
    \left( \int_0^{n_t} [x_t(j)]^{1-1/\sigma} \diff j\right)^\frac{1}{1-1/\sigma}, \label{eq:prod_X}
\end{equation}
where $X_t$ is the output of the knowledge-intensive good, $n_t$ is the measure of intermediate-good varieties existing at date $t$, $x_t(j)$ is the input of the intermediate good of variety $j$, and $\sigma>1$ is the elasticity of substitution across varieties. The coefficient $n_t^{-1/(\sigma-1)}$ in \eqref{eq:prod_X} is a scaling factor to disentangle market power and taste for variety (or productivity): see \citet{Benassy1996,Benassy1998}. For simplicity, we refer to $n_t$ as knowledge. As we describe below, $n_t$ endogenously grows as a result of innovation (R\&D). Taking as given the price of the knowledge-intensive good $P_t>0$ and the prices of intermediate goods $\set{p_t(j)}_{j\in [0,n_t]}$, the knowledge-intensive good firm maximizes the profit
\begin{equation}
    P_tX_t-\int_0^{n_t} p_t(j)x_t(j)\diff j. \label{eq:profit_X}
\end{equation}
Again, because the production function \eqref{eq:prod_X} is homogeneous of degree one, the maximized profit is zero.

\paragraph{Intermediate good sector}

Each intermediate good $j$ is produced by a single monopolistically competitive firm using skilled labor as input. One unit of skilled labor produces one unit of the intermediate good. Thus, taking the wage of skilled labor $w_{Ht}>0$ as given, firm $j$ chooses the price $p_t(j)$ and quantity $x_t(j)$ to maximize the profit
\begin{equation}
    d_t(j)=(p_t(j)-w_{Ht})x_t(j), \label{eq:profit_j}
\end{equation}
where firm $j$ understands that $x_t(j)$ equals the demand obtained by maximizing \eqref{eq:profit_X}. The profit $d_t(j)$ in \eqref{eq:profit_j} is distributed as dividends to the shareholders of firm $j$.

\paragraph{R\&D sector}

New intermediate good varieties are created through R\&D. Letting $n_t$ be knowledge at date $t$, a unit of skilled labor engaging in R\&D at that date creates $an_t$ new intermediate-good varieties, where $a>0$ is a productivity parameter. The term $n_t$ in the R\&D production function $an_t$ captures knowledge spillover from existing innovations to new ones.\footnote{As is well known in the endogenous growth literature such as \citet{Romer1990}, this linearity assumption of knowledge spillover is necessary to generate endogenous growth. Our setup is equivalent to Equation (3.21) of \citet[p.~58]{GrossmanHelpman1991book}, where our $a$ and $n_t$ correspond to their $1/a$ and $K_n$.}

New varieties are protected by patents or kept as trade secrets. Each new firm $j$ issues one share in an initial public offering (IPO) at price $q_t(j)$. The share is a claim to the stream of monopoly profits \eqref{eq:profit_j}, and the IPO proceeds accrue to the founding skilled agent. Thus, $q_t(j)$ can also be interpreted as the price of an idea or a unit of protected knowledge, \ie, intangible capital.

At $t=0$, the initial old own the $n_0$ incumbent firms. As shown below in \eqref{eq:dtj}, symmetry implies that all firms pay the same dividend $D_t\coloneqq d_t(j)$. Because incumbent ex-dividend shares and newly issued shares are identical claims to dividends beginning at $t+1$, they trade at the same ex-dividend price. Throughout, $D_t$ denotes the dividend per firm (and per share), so aggregate dividends are $n_tD_t$.

\subsection{Equilibrium}

Given the initial knowledge $n_0$ and the productivity process $\set{(A_{Xt},A_{Lt})}_{t=0}^\infty$, a rational expectations equilibrium consists of prices and allocations such that agents optimize, firms maximize profits, and all goods, labor, and financial markets clear.

\paragraph{Firms}

The first-order conditions for the profit maximization problem \eqref{eq:profit_Y} are
\begin{subequations}
\begin{align}
    P_t&=F_X(A_{Xt}X_t,A_{Lt}L)A_{Xt}, \label{eq:Pt} \\
    w_{Lt}&=F_L(A_{Xt}X_t,A_{Lt}L)A_{Lt}, \label{eq:wLt}
\end{align}
\end{subequations}
where we have imposed the unskilled labor market clearing condition $L_t=L$. The first-order condition for the knowledge-intensive-good firm yields the demand for intermediate good $j$:
\begin{equation}
    x_t(j)=(X_t/n_t)(p_t(j)/P_t)^{-\sigma} \label{eq:xtj}
\end{equation}
Substituting \eqref{eq:xtj} into \eqref{eq:profit_j} gives the familiar monopoly price
\begin{equation}
    p_t(j)=\frac{\sigma}{\sigma-1}w_{Ht}, \label{eq:ptj}
\end{equation}
which marks up marginal cost $w_{Ht}$ by the factor $\sigma/(\sigma-1)>1$ \citep{DixitStiglitz1977}. Symmetry implies $x_t(j)\eqqcolon x_t$, and hence
\begin{equation}
    X_t=n_t^{-1/(\sigma-1)}
    (n_tx_t^{(\sigma-1)/\sigma})^{\sigma/(\sigma-1)}=n_tx_t. \label{eq:Xt}
\end{equation}
Consequently, $P_t=p_t(j)=\sigma w_{Ht}/(\sigma-1)$, and the dividend is
\begin{equation}
    d_t(j)=\frac{w_{Ht}x_t}{\sigma-1}. \label{eq:dtj}
\end{equation}

\paragraph{Agents}

Imposing the law of one price, as in ordinary macro-finance models, let $Q_t\coloneqq q_t(j)$ denote the common ex-dividend price of all stocks.

A skilled agent can either work in the intermediate-good sector and earn $w_{Ht}$ or engage in R\&D, create $an_t$ new firms, and receive the IPO proceeds $an_tQ_t$. Letting $\phi_t$ denote the fraction of skilled agents who work in intermediate-good production, occupational optimality requires
\begin{equation}
    an_tQ_t\le w_{Ht},\qquad
    (1-\phi_t)(w_{Ht}-an_tQ_t)=0. \label{eq:indifference}
\end{equation}
Thus, if innovation occurs ($\phi_t<1$), skilled agents are indifferent between the two occupations and $an_tQ_t=w_{Ht}$. Using \eqref{eq:Xt}--\eqref{eq:indifference}, the dividend yield in this case is
\begin{equation}
    \frac{D_t}{Q_t}=\frac{w_{Ht}x_t}{\sigma-1}/\frac{w_{Ht}}{an_t}
    =\frac{aX_t}{\sigma-1}. \label{eq:divyield1}
\end{equation}

Let $\theta_{it}$ be the total shares held at the end of date $t$ by an agent of type $i\in \set{H,L}$, and let $(c_{it}^y,c_{i,t+1}^o)$ denote consumption when young and old. The reduced-form budget constraints are
\begin{subequations}\label{eq:budget_GH}
\begin{align}
    &\text{Young:} & c_{it}^y+Q_t\theta_{it}&=w_{it}, \label{eq:budget_GH_young}\\
    &\text{Old:} & c_{i,t+1}^o&=(Q_{t+1}+D_{t+1})\theta_{it}. \label{eq:budget_GH_old}
\end{align}
\end{subequations}
As in \S\ref{sec:toy}, optimal consumption is $c_{it}^y=(1-\beta)w_{it}$ and savings equals $\beta w_{it}$. Hence, in every equilibrium,
\begin{equation}
    \theta_{it}=\frac{\beta w_{it}}{Q_t},\qquad i\in\set{H,L}. \label{eq:stock_demand_GH}
\end{equation}
\paragraph{Market clearing}

The allocation of skilled labor implies
\begin{equation}
    X_t=n_tx_t=\phi_t H. \label{eq:laborH}
\end{equation}
Because the mass of skilled agents conducting R\&D is $(1-\phi_t)H$, end-of-date stock supply evolves according to
\begin{equation}
    n_{t+1}=(1+a(1-\phi_t)H)n_t. \label{eq:n_dyn}
\end{equation}
The stock market clearing condition at the end of period $t$ is
\begin{equation}
    \underbrace{n_{t+1}}_\text{supply}=\underbrace{H\theta_{Ht}+L\theta_{Lt}}_\text{demand}. \label{eq:n_clear}
\end{equation}

Combining occupational choice with labor- and stock-market clearing reduces equilibrium to the determination of $\phi_t$. To state the result compactly, define relative effective productivity by $\zeta_t\coloneqq (A_{Xt}H)/(A_{Lt}L)$.

\begin{prop}[Equilibrium]\label{prop:eq_GH}
Suppose Assumptions \ref{asmp:utility} and \ref{asmp:F} hold. Then there exists a unique rational expectations equilibrium. If
\begin{equation}
    \frac{1}{aH}<\beta+\frac{\beta\sigma}{(\sigma-1)\zeta_tg(\zeta_t)}, \label{eq:phi_cond}
\end{equation}
innovation occurs, and the fraction $\phi_t\in(0,1)$ of skilled agents working in the intermediate-good sector uniquely satisfies
\begin{equation}
    \frac{1}{aH}=\phi_t-1+\beta
    +\frac{\beta\sigma}{(\sigma-1)\zeta_tg(\zeta_t\phi_t)}. \label{eq:eqcond1}
\end{equation}
If \eqref{eq:phi_cond} does not hold, then $\phi_t=1$ and there is no innovation. In either case, $\phi_t$ depends on $t$ only through $\zeta_t$, or equivalently the relative productivity $A_{Xt}/A_{Lt}$. The goods prices and wages are
\begin{subequations}\label{eq:eqprice}
\begin{align}
    &\text{Knowledge-intensive good price:} & P_t&=p_t(j)=F_XA_{Xt}, \label{eq:eqprice_P}\\
    &\text{Skilled wage:} & w_{Ht}&=\frac{\sigma-1}{\sigma}F_XA_{Xt}, \label{eq:eqprice_wH}\\
    &\text{Unskilled wage:} & w_{Lt}&=F_LA_{Lt}, \label{eq:eqprice_wL}
\end{align}
\end{subequations}
where $F_X,F_L$ are evaluated at $(A_{Xt}H\phi_t,A_{Lt}L)$. The stock price is
\begin{equation}
    Q_t=
    \begin{cases*}
        \dfrac{w_{Ht}}{an_t}=\dfrac{\sigma-1}{\sigma an_t}F_XA_{Xt}, & if \eqref{eq:phi_cond} holds,\\
        \dfrac{\beta(Hw_{Ht}+Lw_{Lt})}{n_t}, & otherwise.
    \end{cases*} \label{eq:eqprice_Q}
\end{equation}
\end{prop}

\section{Emergence of stochastic stock price bubbles}\label{sec:GH_bubble}

This section derives our main results. We first specialize the productivity process and characterize innovation and knowledge growth. We then establish bubble necessity, examine the role of bubble-backed IPO proceeds and the consequences of the transition to balanced growth, and explain why balanced growth requires knife-edge restrictions. We conclude by showing that finite lifetimes are not essential for the bubble result.

\subsection{Technology regimes and knowledge growth}

Following the setting in \S\ref{sec:toy}, agents have Epstein-Zin utility (Assumption \ref{asmp:utility}). The aggregate state is $z_t\in \set{u,b}$, where $u$ and $b$ correspond to unbalanced and balanced growth, respectively, and $b$ is absorbing (Assumption \ref{asmp:states}). The production function is \eqref{eq:F_CES}. We impose the following assumption on factor-augmenting productivities.

\begin{asmp}\label{asmp:AZ}
There exist constants $A_X,A_L>0$ and $\xi_u,\xi_b,\lambda_u,\lambda_b\ge 0$ such that
\begin{equation}
    (A_{Xt},A_{Lt})=(A_Xn_t^{\xi_{z_t}},A_Ln_t^{\lambda_{z_t}}). \label{eq:AXLt}
\end{equation}
Furthermore, the following conditions hold:
\begin{equation}
    \kappa\coloneqq (\xi_u-\lambda_u)(\rho-1)>0 \quad \text{and} \quad
    \lambda_u>\lambda_b=\xi_b. \label{eq:xi}
\end{equation}
\end{asmp}

This specialization maps the three features of GPTs characterized by \citet{BresnahanTrajtenberg1995} (Footnote \ref{fn:GPT}) into our model as follows:
\begin{enumerate*}
    \item pervasiveness is captured by \eqref{eq:AXLt}, in which knowledge affects both the productivity of the knowledge-intensive input, $A_{Xt}$, and the productivity of unskilled labor, $A_{Lt}$;
    \item improvement is represented by the accumulation of knowledge $n_t$, which expands the stock of available product varieties; and
    \item innovation spanning arises because the existing stock of knowledge raises R\&D productivity: one unit of skilled labor devoted to R\&D creates $an_t$ new varieties.
\end{enumerate*}
The transition from state $u$ to state $b$ provides a reduced-form representation of the technology's maturation and diffusion: the exceptional and uneven spillovers associated with an emerging technology subside, and further knowledge accumulation affects both factor productivities at the same rate. When $\rho>1$, this common rate is weaker than either state-$u$ spillover. We do not explicitly describe adoption choices or diffusion across firms and sectors, or endogenize the transition between technology regimes; consequently, we abstract from forces determining when the technology matures or when the associated stock price bubble collapses.
This interpretation is motivated by evidence that GPTs diffuse through gradual complementary investment and organizational change \citep{JovanovicRousseau2005,BrynjolfssonRockSyverson2021}. AI may also raise R\&D productivity by serving as a general-purpose method of invention \citep{CockburnHendersonStern2019}, consistent with our innovation-spanning channel.

Knowledge spillovers to both production factors are common in endogenous growth models \citep{Frankel1962,Romer1986}. The first condition in \eqref{eq:xi} makes these spillovers unbalanced in state $u$: as knowledge accumulates, the productivities of the two factors grow at different rates. In the empirically relevant case in which their elasticity of substitution $1/\rho$ is below one \citep{OberfieldRaval2021,GechertHavranekIrsovaKolcunova2022}, this condition corresponds to $\xi_u>\lambda_u$, so new knowledge initially benefits the knowledge-intensive sector more strongly. For example, a frontier AI model may first raise the productivity of researchers, programmers, and other knowledge-intensive workers. As user-friendly applications and complementary business practices diffuse---as occurred with personal computers, tablets, the Internet, and AI assistants---the same technology also makes routine work more efficient. The restriction $\lambda_b=\xi_b$ implies that the economy returns after the transition to a balanced growth path on which aggregate consumption, R\&D investment, and dividends are constant fractions of GDP. This equality is a knife-edge restriction, as is generally required for balanced growth (\S\ref{sec:uzawa}).

Under the maintained assumptions, whenever innovation occurs, the equilibrium condition \eqref{eq:eqcond1} reduces to
\begin{equation}
    \frac{1}{aH}=\phi_t-1+\beta
    +\frac{\beta\sigma(1-\alpha)}{(\sigma-1)\alpha}
    \left(\frac{A_XH}{A_LL}\right)^{\rho-1}
    n_t^{(\xi_{z_t}-\lambda_{z_t})(\rho-1)}\phi_t^\rho. \label{eq:eqcond2}
\end{equation}
For results requiring innovation throughout state $u$, we impose the initial condition
\begin{equation}
    \frac{1}{aH}<\beta+\frac{\beta\sigma(1-\alpha)}{(\sigma-1)\alpha}
    \left(\frac{A_XH}{A_LL}\right)^{\rho-1}
    n_0^{(\xi_u-\lambda_u)(\rho-1)}. \label{eq:n0}
\end{equation}
Noting that \eqref{eq:xi} holds, \eqref{eq:n0} implies that $n_0$ is sufficiently large. Condition \eqref{eq:n0} ensures that condition \eqref{eq:phi_cond} holds at $t=0$. Because knowledge grows while state $u$ persists, it is necessary and sufficient for innovation to occur at every date before the transition to state $b$.

The following proposition characterizes the asymptotic gross growth factor of knowledge. To state the result, we define $\phi_b\in (0,1]$ as follows. If
\begin{equation}
    \frac{1}{aH}>\beta+\frac{\beta\sigma(1-\alpha)}{(\sigma-1)\alpha}
    \left(\frac{A_XH}{A_LL}\right)^{\rho-1}, \label{eq:phib_cond1}
\end{equation}
we define $\phi_b=1$. If \eqref{eq:phib_cond1} does not hold, we define $\phi_b\in (0,1]$ by
\begin{equation}
    \frac{1}{aH}=\phi_b-1+\beta
    +\frac{\beta\sigma(1-\alpha)}{(\sigma-1)\alpha}
    \left(\frac{A_XH}{A_LL}\right)^{\rho-1}\phi_b^\rho. \label{eq:phib_cond2}
\end{equation}

\begin{prop}[Innovation and knowledge growth]\label{prop:growth}
Suppose the production function is \eqref{eq:F_CES}, Assumptions \ref{asmp:utility}, \ref{asmp:states}, and \ref{asmp:AZ} hold, and $n_0$ satisfies \eqref{eq:n0}. Then the following statements are true.
\begin{enumerate}
    \item\label{item:growth_u} Conditional on staying in state $u$, $\phi_t$ decreases monotonically to zero and $n_{t+1}/n_t\to G_u\coloneqq 1+aH$.
    \item\label{item:growth_b} In state $b$, $\phi_t=\phi_b$ and $n_{t+1}/n_t=G_b\coloneqq 1+a(1-\phi_b)H<G_u$.
\end{enumerate}
\end{prop}

The intuition for Proposition \ref{prop:growth} is as follows. Because $(\xi_u-\lambda_u)(\rho-1)>0$ by \eqref{eq:xi}, in state $u$ the right-hand side of \eqref{eq:eqcond2} is increasing in both $n_t$ and $\phi_t$. Therefore, as long as state $u$ persists, $n_t$ increases and $\phi_t$ decreases monotonically to 0. By \eqref{eq:n_dyn}, $n_{t+1}/n_t\to G_u=1+aH$. In other words, the share of skilled labor devoted to R\&D, $1-\phi_t$, increases monotonically, expanding product variety and accelerating innovation. Once the economy switches to state $b$, the occupational condition no longer depends on $n_t$ because $\lambda_b=\xi_b$. Consequently, $\phi_t$ remains constant at $\phi_b$. When the reverse of \eqref{eq:phib_cond1} is strict, \eqref{eq:phib_cond2} yields $\phi_b<1$ and innovation continues; at equality or when \eqref{eq:phib_cond1} holds, $\phi_b=1$ and innovation stops. In either case, R\&D activity and knowledge growth fall relative to state $u$.

\begin{figure}[!htb]
    \centering
    \includegraphics[width=0.7\linewidth]{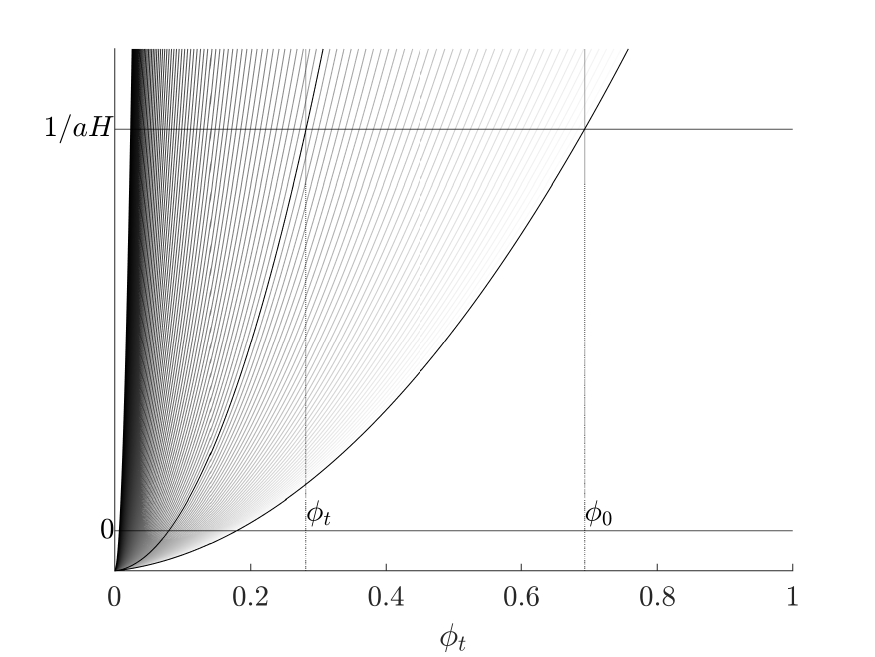}
    \caption{Determination of equilibrium $\phi_t$.}
    \label{fig:GH_phi}
    \caption*{\footnotesize Note: The numerical example is based on parameter values $\beta=1/2$, $\alpha=1/2$, $\rho=2$, $\sigma=2$, $A_XH=10$, $A_LL=1$, $\xi_u=0.7$, $\lambda_u=0.2$, $\xi_b=\lambda_b=0.1$, $aH=0.2$, and $n_0=1$.}
\end{figure}

Figure \ref{fig:GH_phi} shows how $\phi_t$ is determined in equilibrium, assuming state $u$ persists. The upward-sloping curves are the right-hand side of \eqref{eq:eqcond2}. The horizontal line is the level $1/aH$, which is the left-hand side of \eqref{eq:eqcond2}. The equilibrium $\phi_t$ is determined as the intersection of the two graphs. As time passes, the upward-sloping curves become steeper, so $\phi_t$ decreases.

\subsection{Bubble necessity}

We are now ready to show the emergence of stochastic bubbles. For either type $i\in\set{H,L}$, the consumption and portfolio rules imply
\begin{equation*}
    c_{it}^y=(1-\beta)w_{it},\qquad
    c_{i,t+1}^o=\frac{Q_{t+1}+D_{t+1}}{Q_t}\beta w_{it}.
\end{equation*}
Substituting these expressions into the SDF \eqref{eq:SDF}, the wage $w_{it}$ cancels and we obtain exactly the same SDF \eqref{eq:SDF_Q} as in \S\ref{sec:toy}. This derivation applies whether or not innovation occurs. Therefore, to determine the existence or nonexistence of bubbles, it suffices to check the assumptions of Theorem \ref{thm:bubble_toy}.

Using \eqref{eq:dtj}, \eqref{eq:laborH}, and the piecewise stock price \eqref{eq:eqprice_Q}, we obtain the dividend yield
\begin{equation}
    \frac{D_t}{Q_t}=
    \begin{cases*}
        \dfrac{aH}{\sigma-1}\phi_t, & if innovation occurs,\\
        \dfrac{1}{\beta(\sigma-1)}\dfrac{Hw_{Ht}}{Hw_{Ht}+Lw_{Lt}}, & otherwise.
    \end{cases*} \label{eq:divyield2}
\end{equation}
Once state $b$ is reached, $\phi_t$ is constant at $\phi_b$ by Proposition \ref{prop:growth}. If innovation continues, the per-share price and dividend change by the common gross factor $G_b^{\lambda_b-1}$; if innovation stops, both are constant. Thus the dividend yield $D_t/Q_t$ is a positive constant. Because all uncertainty is resolved after the transition, the SDF reduces to $m_{t\to t+1}=Q_t/(Q_{t+1}+D_{t+1})$. The same argument as in Proposition \ref{prop:nobubble_b} then implies $Q_t=V_t$.

In state $u$, Proposition \ref{prop:growth}, the equilibrium condition \eqref{eq:eqcond2}, and the dividend-yield formula \eqref{eq:divyield2} imply that the price-dividend ratio asymptotically grows exponentially. Upon a transition to state $b$, the relative stock price $Q_t^b/Q_t^u$ decreases with accumulated knowledge because $\lambda_u>\lambda_b$. Together with $\gamma<1$, these two properties verify the conditions of Theorem \ref{thm:bubble_toy}. Consequently, we obtain our main result.

\begin{thm}[Bubble necessity with innovation]\label{thm:bubble_GH}
Suppose the production function is \eqref{eq:F_CES}, Assumptions \ref{asmp:utility}, \ref{asmp:states}, and \ref{asmp:AZ} hold, $n_0$ satisfies \eqref{eq:n0}, and $\gamma<1$. Let $Q_t$ be the stock price in the unique equilibrium established in Proposition \ref{prop:eq_GH} and $V_t$ be its fundamental value. Then the following statements are true.
\begin{enumerate}
    \item If $z_t=u$, the stock price exhibits a bubble: $Q_t>V_t$. Conditional on remaining in state $u$, the price-dividend ratio $Q_t/D_t$ asymptotically grows exponentially.
    \item If $z_t=b$, the stock price reflects fundamentals: $Q_t=V_t$ and the price-dividend ratio $Q_t/D_t$ is constant. 
\end{enumerate}
\end{thm}

Theorem \ref{thm:bubble_GH} translates the two bubble conditions of Theorem \ref{thm:bubble_toy} into restrictions on technology. The condition $\kappa=(\xi_u-\lambda_u)(\rho-1)>0$ makes the dividend yield vanish and the price-dividend ratio grow exponentially while state $u$ persists, whereas $\lambda_u>\lambda_b=\xi_b$ makes the transition-state stock price small relative to the continuation-state price: $Q_t^b/Q_t^u\sim n_t^{\lambda_b-\lambda_u}$. Together with $\gamma<1$, these forces generate a bubble even though agents anticipate its eventual collapse and the return to balanced growth. The reduced-form lag in economy-wide spillovers connects the model to the time-lag explanation of the Solow productivity paradox \citep{Brynjolfsson1993}, while the exponentially growing price-dividend ratio connects it to the econometric literature on bubble detection \citep*{PhillipsShiYu2015,PhillipsShi2018,PhillipsShi2020}.

\begin{figure}[!htb]
    \centering
    \includegraphics[width=0.7\linewidth]{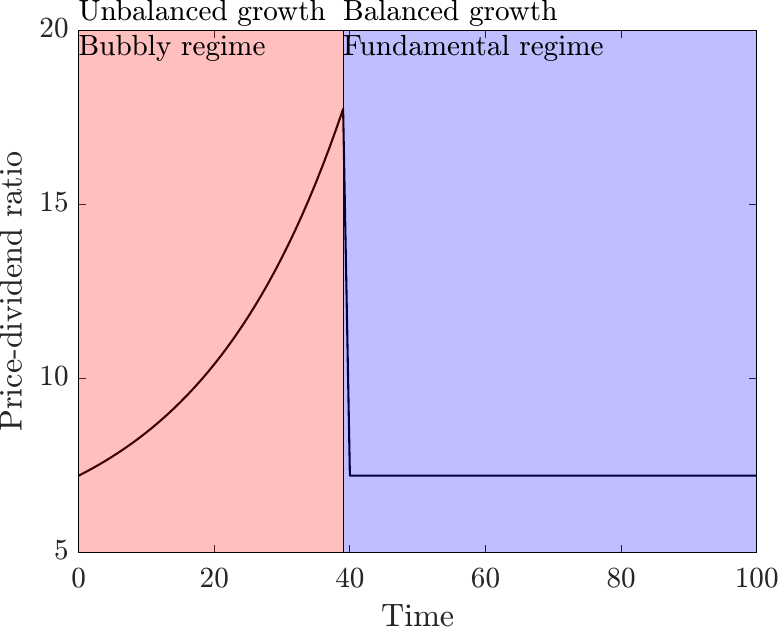}
    \caption{Dynamics of price-dividend ratio.}
    \label{fig:GH_PD}
\end{figure}

Figure \ref{fig:GH_PD} shows the dynamics of the price-dividend ratio for the same numerical example as in Figure \ref{fig:GH_phi}, computed as the reciprocal of the dividend yield \eqref{eq:divyield2}. Here, we suppose that the state $u$ persists until $t=39$ but switches to $b$ at $t=40$. During the bubble, the price-dividend ratio grows exponentially. Once the bubble bursts, the price-dividend ratio becomes constant (because $\phi_b$ is constant) and the economy switches to balanced growth.

\citet[p.~27]{Ueda1990} made a closely related observation about Japanese equities in the late 1980s (translated from Japanese):
\begin{quote}
	Japanese stock prices stand at abnormally high levels relative to corporate fundamentals such as earnings and dividends. [\ldots] If a bubble emerges in stock prices while underlying corporate conditions show little improvement, the dividend yield will converge to zero, and capital gains will come to account for nearly all returns on stock investment. Thus, recent stock-price movements are also consistent with the view that they constitute a bubble.
\end{quote}
Indeed, when the dividend yield converges to zero, we can write the net stock return as
\begin{equation*}
    \frac{Q_t+D_t}{Q_{t-1}}-1=\underbrace{\frac{Q_t-Q_{t-1}}{Q_{t-1}}}_\text{capital gain}+\frac{Q_t}{Q_{t-1}}\times\underbrace{\frac{D_t}{Q_t}}_{\text{dividend yield}\approx 0},
\end{equation*}
so stock returns asymptotically consist solely of capital gains from speculative resale.

\subsection{Bubble-backed innovation and regime transition}\label{sec:macro_consequences}

The equilibrium path in state $u$ combines unusually strong and uneven productivity spillovers with a stock price bubble. Comparisons across histories therefore do not isolate the role of the bubble itself. We first use a partial-equilibrium experiment to identify the direct effect of capitalizing the bubble component into the proceeds from innovative activity. We then return to the full equilibrium and characterize the consequences of the transition to state $b$.
This experiment is motivated by evidence that overvalued equity can stimulate issuance and real investment \citep{GilchristHimmelbergHuberman2005} and that external equity is an important source of R\&D finance for young high-technology firms \citep{BrownFazzariPetersen2009}.

Using \eqref{eq:prod_Y}, \eqref{eq:F_CES}, \eqref{eq:laborH}, and \eqref{eq:AXLt}, equilibrium output is
\begin{equation}
    Y_t=
    \left[
    \alpha(A_Xn_t^{\xi_{z_t}}\phi_tH)^{1-\rho}
    +(1-\alpha)(A_Ln_t^{\lambda_{z_t}}L)^{1-\rho}
    \right]^{\frac{1}{1-\rho}}. \label{eq:Yt_CES}
\end{equation}

Fix a date $t$ and a history in state $u$. Decompose the equilibrium stock price as $Q_t=V_t+B_t$, as in \eqref{eq:decompose}. For $\chi\in[0,1]$, suppose that a skilled agent who creates $an_t$ firms receives net IPO proceeds $an_tQ_t^{\mathrm{IPO}}(\chi)$, where
\begin{equation}
    Q_t^{\mathrm{IPO}}(\chi)\coloneqq V_t+\chi B_t. \label{eq:IPO_capture}
\end{equation}
The parameter $\chi$ is the fraction of the bubble component captured by the innovator. All shares continue to trade at the common price $Q_t$ under the law of one price; $Q_t^{\mathrm{IPO}}(\chi)$ denotes the net proceeds per new share retained by the innovator, not a distinct transaction price. The benchmark equilibrium corresponds to $\chi=1$, whereas $\chi=0$ assigns the innovator only the fundamental (dividend-backed) component $V_t$. The residual bubble component may be interpreted as accruing to an outside financial sector or as belonging to a separate pure bubble asset.

For $\phi\in(0,1]$, define contemporaneous output and wages by
\begin{subequations}\label{eq:PE_outcomes}
\begin{align}
    Y_t(\phi)&\coloneqq F(A_{Xt}H\phi,A_{Lt}L),\\
    w_{Ht}(\phi)&\coloneqq \frac{\sigma-1}{\sigma}A_{Xt}F_X(A_{Xt}H\phi,A_{Lt}L),\\
    w_{Lt}(\phi)&\coloneqq A_{Lt}F_L(A_{Xt}H\phi,A_{Lt}L).
\end{align}
\end{subequations}
Let $\phi_t(\chi)\in(0,1]$ satisfy the occupational choice conditions
\begin{equation}
    an_tQ_t^{\mathrm{IPO}}(\chi)\le w_{Ht}(\phi_t(\chi)),
    \quad
    (1-\phi_t(\chi))
    \left[w_{Ht}(\phi_t(\chi))-an_tQ_t^{\mathrm{IPO}}(\chi)\right]=0,
    \label{eq:PE_occupation}
\end{equation}
and define the implied next-period knowledge stock by
\begin{equation}
    \widetilde n_{t+1}(\chi)
    \coloneqq
    \left[1+aH(1-\phi_t(\chi))\right]n_t. \label{eq:PE_knowledge}
\end{equation}

\begin{prop}[Bubble-backed IPO proceeds]\label{prop:IPO_capture}
Suppose the conditions of Theorem \ref{thm:bubble_GH} hold, and fix a date $t$ on an equilibrium history in state $u$. Then $\phi_t(\chi)$ is uniquely determined by \eqref{eq:PE_occupation} and decreases with $\chi$. Consequently, the skilled wage $w_{Ht}$, the skilled-to-unskilled wage ratio $w_{Ht}/w_{Lt}$, R\&D employment $H(1-\phi_t)$, and next-period knowledge $\widetilde n_{t+1}$ increase with $\chi$, whereas the unskilled wage $w_{Lt}$ and current output $Y_t$ decrease with $\chi$.
\end{prop}

Proposition \ref{prop:IPO_capture} identifies the direct effect of attaching the bubble to productive equity. When innovators capture the bubble component, founding a firm becomes more attractive. Skilled labor moves from current production into R\&D, raising the marginal product and wage of the skilled agents who remain in production. The same reallocation lowers current output and the unskilled wage but expands product variety available from the next date onward. Under CES production, the wage-ratio result follows transparently from $w_{Lt}(\phi)/w_{Ht}(\phi)\propto\phi^\rho$. It is consistent with the widening wage gap during the IT wave documented by \citet[Figure~18]{JovanovicRousseau2005} and \citet[\S4]{VanReenen2011}.

Because the benchmark values of $n_t$, $A_{Xt}$, $A_{Lt}$, $V_t$, and $B_t$ are held fixed and financial-market clearing and wealth effects are suppressed, the proposition is a one-period partial-equilibrium incidence experiment, not an alternative rational expectations equilibrium. In particular, $V_t$ remains the benchmark fundamental component, and \eqref{eq:PE_knowledge} cannot be iterated without recomputing prices, dividends, and the SDF.

We next return to the model's equilibrium path. The technology regime and the valuation regime change simultaneously when the state switches from $u$ to $b$, so the next result concerns their joint transition rather than the causal effect of the bubble holding technology fixed.

\begin{prop}[Consequences of the regime transition]\label{prop:transition_consequences}
Suppose the conditions of Theorem \ref{thm:bubble_GH} hold.
\begin{enumerate}
    \item\label{item:transition1} The ratios of output and the per-share stock price across the transition satisfy
    \begin{equation}
        \frac{Y_t^b}{Y_{t-1}^u}\sim n_t^{\lambda_b-\lambda_u}\to0,
        \qquad
        \frac{Q_t^b}{Q_{t-1}^u}\sim n_t^{\lambda_b-\lambda_u}\to0
        \quad\text{as }t\to\infty. \label{eq:transition_declines}
    \end{equation}
    \item\label{item:transition2} Let $\tau$ denote the date of the first transition to state $b$. There exist constants $C_Y,C_H,C_L>0$, independent of $\tau$ and $k$, such that, for every $k\ge0$,
    \begin{subequations}\label{eq:post_transition_levels}
    \begin{align}
        Y_{\tau+k}&=C_YG_b^{\lambda_b k}n_\tau^{\lambda_b},\\
        w_{H,\tau+k}&=C_HG_b^{\lambda_b k}n_\tau^{\lambda_b},\\
        w_{L,\tau+k}&=C_LG_b^{\lambda_b k}n_\tau^{\lambda_b}.
    \end{align}
    \end{subequations}
    Hence, at every fixed horizon after the transition, a later transition increases output and both wages through the additional accumulation of knowledge. The post-transition wage ratio is independent of the transition date.
\end{enumerate}
\end{prop}

The first part of Proposition \ref{prop:transition_consequences} shows that output and the per-share stock price fall at the same asymptotic order, $n_t^{-(\lambda_u-\lambda_b)}$. It does not imply that the two percentage declines are equal: their proportionality constants depend on parameters and, for the stock price, on whether innovation continues in state $b$. These declines reflect the growing difference between the unbalanced- and balanced-growth regimes and should not be interpreted as the pure effect of the bubble's collapse. The second part provides an exact propagation result: any additional knowledge inherited at the transition permanently raises the level paths of output and wages when $\lambda_b>0$, without changing their balanced-growth rates or the skill premium. Together with Proposition \ref{prop:IPO_capture}, it identifies a productive-legacy channel through which bubble-backed IPO proceeds can matter beyond the bubble phase (Footnote \ref{fn:Bezos}). This conclusion concerns productive capacity, not welfare across generations.

Figure \ref{fig:GH_Y} illustrates the equilibrium implications of early and late transitions. In the numerical example, the state switches to $b$ at $\tau=30$ or $\tau=50$. A later transition produces a larger contemporaneous contraction but leaves the economy with a larger stock of knowledge and hence a higher post-transition output path.

\begin{center}
    \centering
    \includegraphics[width=0.7\linewidth]{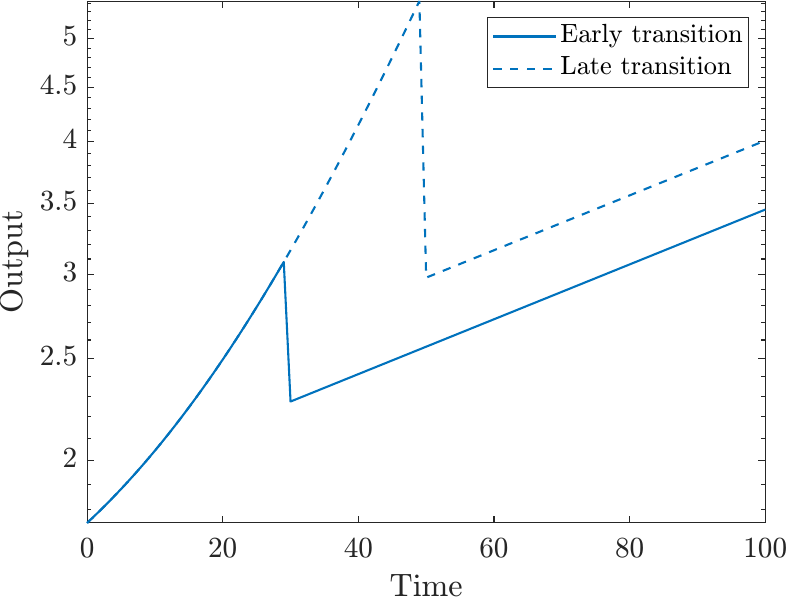}
    \captionsetup{hypcap=false}
    \captionof{figure}{Output under early and late transitions to balanced growth.}
    \label{fig:GH_Y}
\end{center}

\subsection{Knife-edge restrictions and macro-financial modeling}\label{sec:uzawa}

Balanced growth is a convenient device and a standard benchmark in macroeconomic modeling. Yet it generally does not arise under arbitrary technologies and patterns of technical change. Unbalanced growth may take other forms: \citet{AcemogluGuerrieri2008}, for example, obtain sectoral reallocation while aggregate behavior remains consistent with the Kaldor facts. Here the relevant imbalance is in factor-augmenting productivity, which affects the aggregate price-dividend ratio. Macroeconomic models are often parameterized---most prominently through Cobb--Douglas production or restrictions on the direction of technical change---so that the relevant aggregates share a common growth trend. In our model, the asset-pricing counterpart of balanced growth is a constant price-dividend ratio: stock prices and dividends grow at the same rate. Once this property is built into the model, the relative-growth conditions in Theorem \ref{thm:bubble_GH} cannot hold. The relevant question is therefore whether a constant price-dividend ratio is a robust implication or a consequence of special restrictions. The next proposition provides a precise answer.

\begin{prop}[Balanced growth is knife-edge]\label{prop:knife-edge}
Consider the economy of \S\ref{subsec:innovation_model}, and suppose Assumptions \ref{asmp:utility} and \ref{asmp:F} hold and innovation occurs for every possible value of the relative productivity $A_{Xt}/A_{Lt}>0$. Then the equilibrium price-dividend ratio is invariant to relative productivity if and only if the production function is Cobb--Douglas. In particular, if the production function is \eqref{eq:F_CES} and productivities take the form $(A_{Xt},A_{Lt})=(A_Xn_t^\xi,A_Ln_t^\lambda)$ for two consecutive dates, the price-dividend ratio is constant if and only if
\begin{equation}
    (\xi-\lambda)(\rho-1)=0. \label{eq:knife-edge}
\end{equation}
\end{prop}
Proposition \ref{prop:knife-edge} shows that a constant price-dividend ratio is knife-edge in two precise senses. Across all possible values of relative productivity, invariance of the price-dividend ratio characterizes the Cobb--Douglas production function. Within the CES class, the ratio can remain constant along a growing path only if the two factor-augmenting productivities grow at the same rate, $\xi=\lambda$, or the production function is Cobb--Douglas, $\rho=1$. Both are equality restrictions. Outside these cases, knowledge accumulation changes relative productivity and necessarily changes the price-dividend ratio.

This result is the asset-pricing analogue of the Uzawa steady-state growth theorem. \citet[p.~118]{Uzawa1961} defined technical change to be Harrod-neutral when a constant capital-output ratio implies a constant marginal product of capital, and showed that this property requires a representation of the form $F(K,A(t)L)$. The theorem subsequently came to be understood as stating that balanced growth in the neoclassical growth model requires technical change to be effectively labor-augmenting \citep{JonesScrimgeour2008}. \citet{Schlicht2006} established a related result: if output, capital, and consumption all grow exponentially, possibly at different constant rates, then the production function must again take the form $F(K,A(t)L)$ with exponentially growing $A(t)$. \citet{CaseyHorii2024} generalize the theorem to multifactor production: balanced growth can accommodate capital-augmenting technical change if capital has unit elasticity of substitution with at least one other factor. This result broadens the admissible technologies but retains an equality restriction on substitution. Directed technical change or endogenous schooling may generate the required balance endogenously, but these mechanisms likewise impose restrictions that preserve it \citep[\S15.6]{Acemoglu2009}; see also \citet{GrossmanHelpmanOberfieldSampson2017}. Cobb--Douglas production appears to escape restrictions on factor augmentation only because
\begin{equation*}
    F(A_KK,A_LL)=(A_KK)^\alpha(A_LL)^{1-\alpha}
    =K^\alpha(AL)^{1-\alpha}
\end{equation*}
with $A\coloneqq A_K^\frac{\alpha}{1-\alpha}A_L$, so any factor-augmenting productivities can be rewritten as purely labor-augmenting. The Cobb--Douglas specification itself, however, is a knife-edge case within the CES class.

Imposing balanced growth therefore does more than simplify the dynamics: in our model, it selects parameter configurations under which stock prices and dividends share a common trend. Once unbalanced growth is admitted, the strict inequalities in Theorem \ref{thm:bubble_GH} make the price-dividend ratio grow exponentially and the transition-state stock price fall sufficiently sharply, so a bubble becomes necessary despite its anticipated collapse. A related relative-growth argument appears in \citet{HiranoToda2025PNAS}, where land overvaluation necessarily arises when land productivity growth is stagnant and production factors are sufficiently substitutable. We do not claim that every unbalanced growth path supports a bubble; rather, restricting attention to balanced growth can eliminate the relevant conditions by model construction even before the asset pricing question is posed.

\subsection{Robustness to infinitely lived agents}\label{sec:infinitely-lived}

The two-period OLG structure makes the conditions for the bubble transparent, but finite lifetimes are not essential. To show this, we use the demographic structure of \citet{Weil1989JPubE}: every cohort lives forever, but new cohorts continually enter. At each date, continua of skilled and unskilled agents of masses $H$ and $L$ are born. An agent supplies labor and receives a wage only at birth, then saves and consumes out of financial wealth. A legacy cohort owns the $n_0$ firms outstanding at the beginning of date 0. The production structure, innovation technology, and regime process are otherwise unchanged.

We replace the utility specification in Assumption \ref{asmp:utility} with the following recursive preferences.

\begin{asmp}\label{asmp:utility_BY}
Let $\beta\in(0,1)$ and $\gamma<1$. An agent of type $i\in\set{H,L}$ born at date $\tau$ has recursive utility, at every date $t\ge\tau$,
\begin{equation}
    U_{i,\tau,t}
    =c_{i,\tau,t}^{1-\beta}
    \E_t[U_{i,\tau,t+1}^{1-\gamma}]^{\frac{\beta}{1-\gamma}}. \label{eq:utility_BY}
\end{equation}
\end{asmp}

The unit-EIS aggregator implies that every agent consumes the fraction $1-\beta$ of wealth and saves the fraction $\beta$ (Footnote \ref{fn:log_consumption}). Homotheticity therefore gives the value function $U_{i,\tau,t}=\mu_tW_{i,\tau,t}$, where the continuation-value coefficient $\mu_t>0$ is common across types and cohorts. Appendix \ref{app:proof_bubble_BY} derives the recursion for this coefficient, the associated SDF, and the aggregate market-clearing equation.

For the CES production function, define
\begin{equation}
    K\coloneqq\frac{\beta\sigma(1-\alpha)}{(\sigma-1)\alpha}
    \left(\frac{A_XH}{A_LL}\right)^{\rho-1},
    \quad
    \eta\coloneqq1+\frac{\beta}{\sigma-1},
    \quad
    M\coloneqq(1-\beta)\left(1+\frac{1}{aH}\right).
    \label{eq:constants_BY}
\end{equation}
To ensure active innovation after every possible history, we impose the conditions
\begin{equation}
    Kn_0^\kappa+\eta>M
    \qquad\text{and}\qquad
    K+\eta>M. \label{eq:active_BY}
\end{equation}
We say that a positive continuation-value process $\set{\mu_t}_{t=0}^\infty$ is \emph{admissible} if, along a history that remains in state $u$,
\begin{equation}
    0<\inf_t(\mu_t^u)^{1-\gamma}
    \le\sup_t(\mu_t^u)^{1-\gamma}<\infty.
    \label{eq:admissible_BY}
\end{equation}
This condition rules out explosive continuation-value solutions.

The following theorem extends Theorem \ref{thm:bubble_GH} to infinitely lived agents.

\begin{thm}[Bubble necessity with infinitely lived agents]\label{thm:bubble_BY}
Suppose the production function is \eqref{eq:F_CES}, Assumptions \ref{asmp:states}, \ref{asmp:AZ}, and \ref{asmp:utility_BY} hold, and \eqref{eq:active_BY} is satisfied. Then the aggregate equilibrium allocation, stock price, admissible continuation-value solution, and associated SDF are unique. Moreover, the following statements hold:
\begin{enumerate}
    \item In state $u$, the stock price contains a bubble, $Q_t>V_t$, and the price-dividend ratio asymptotically grows exponentially conditional on remaining in state $u$.
    \item In state $b$, the stock price reflects fundamentals, $Q_t=V_t$, and the price-dividend ratio is constant.
\end{enumerate}
\end{thm}

Thus, finite lifetimes are not essential, but continual cohort entry remains central. Existing cohorts gradually run down their stock positions, and every agent satisfies the transversality condition, while new cohorts replenish aggregate stock demand by investing part of their labor income. Appendix \ref{app:proof_bubble_BY} provides the formal argument and relates this cohort-entry channel to the portfolio reduction condition of \citet{Kamihigashi2018}.

\section{Concluding remarks}\label{sec:conclusion} 

We show that a rational bubble on a dividend-paying stock can be the unique equilibrium outcome even when agents anticipate its complete collapse. During the unbalanced-growth regime, unequal GPT spillovers cause stock prices to outgrow dividends and the price-dividend ratio to rise exponentially. When the spillovers equalize and the economy transitions to balanced growth, the bubble collapses. Although the transition causes a sharp contraction, a later transition leaves a larger stock of knowledge and therefore higher output and wages at any fixed post-transition horizon. Our IPO experiment illustrates how capitalizing the bubble into innovators' proceeds can reinforce this productive legacy by directing skilled labor toward R\&D.

The broader lesson is that balanced growth is not innocuous for asset pricing. In our model, the restrictions supporting balanced growth also keep the price-dividend ratio constant and thereby eliminate the relative-growth conditions generating the bubble. Allowing temporary unbalanced growth reveals equilibrium asset-price dynamics that a balanced-growth model rules out by construction. This result does not require finite lifetimes, although continual cohort entry remains essential. Endogenizing GPT adoption, diffusion, and the transition that precipitates the collapse are natural directions for future work.

\appendix

\section{Proofs}\label{sec:proof}

\subsection{Proof of Proposition \ref{prop:nobubble_b}}

Without loss of generality, assume $z_0=b$. Because uncertainty is resolved, the stochastic discount factor \eqref{eq:SDF_Q} reduces to $m_{t\to t+1}=Q_t/(Q_{t+1}+D_{t+1})$. Therefore,
\begin{equation}
    m_{0\to t}Q_t=\frac{Q_0}{Q_1+D_1}\times\dots\times \frac{Q_{t-1}}{Q_t+D_t}\times Q_t=Q_0\prod_{s=1}^t\frac{Q_s}{Q_s+D_s}. \label{eq:mQ_b}
\end{equation}
By Assumption \ref{asmp:balanced}, $D_t/e_t=D_{t+1}/e_{t+1}$ is a possibly history-dependent constant. Since $Q_t=\beta e_t$ by Proposition \ref{prop:eq_toy}, the dividend yield $D_t/Q_t$ is also constant. Thus $Q_s/(Q_s+D_s)\in(0,1)$ is constant. Therefore, $\lim_{t\to\infty}\E_0[m_{0\to t}Q_t]=0$, and the no-bubble condition \eqref{eq:nobubble} holds. \hfill \qedsymbol

\subsection{Proof of Proposition \ref{prop:nobubble_toy}}

Let $\widehat Q_t\coloneqq m_{0\to t}Q_t$, where $\widehat Q_0=Q_0$, and define the discounted value associated with remaining in state $u$ through date $t$ by $S_t\coloneqq \pi^t\E_0[\widehat Q_t\mid \tau>t]$. For $j\ge 1$, define the discounted value at the date of transition to state $b$ by
\begin{equation*}
    C_j\coloneqq \pi^{j-1}(1-\pi)\E_0[\widehat Q_j\mid \tau=j].
\end{equation*}
Conditional on $\tau\ge j$, the no-arbitrage condition \eqref{eq:noarbitrage} and the partition between $\tau>j$ and $\tau=j$ imply
\begin{align*}
    S_{j-1}
    &=\pi^j\E_0[\widehat Q_j+m_{0\to j}D_j\mid \tau>j]+\pi^{j-1}(1-\pi)\E_0[\widehat Q_j+m_{0\to j}D_j\mid \tau=j] \\
    &\ge S_j+C_j.
\end{align*}
It follows that $S_t$ is decreasing and $\sum_{j=1}^\infty C_j\le S_0=Q_0$.

Fix $j\ge 1$. On the event $\tau=j$, the same derivation as in \eqref{eq:mQ_b} yields
\begin{equation}
    \widehat Q_t=\widehat Q_j\prod_{s=j+1}^t\frac{Q_s}{Q_s+D_s} \label{eq:mQ_j_new}
\end{equation}
for $t\ge j$. By Assumption \ref{asmp:balanced} and Proposition \ref{prop:eq_toy}, conditional on $\tau=j$, the ratio $Q_s/(Q_s+D_s)$ is a constant in $(0,1)$ for $s\ge j$. Therefore, $\widehat Q_t\to 0$ and $0\le \widehat Q_t\le \widehat Q_j$ on the event $\tau=j$. For $j,t\ge 1$, set
\begin{equation*}
    C_{jt}\coloneqq
    \begin{cases*}
        \pi^{j-1}(1-\pi)\E_0[\widehat Q_t\mid \tau=j] & if $j\le t$,\\
        0 & if $j>t$.
    \end{cases*}
\end{equation*}
For each fixed $j$, \eqref{eq:mQ_j_new} implies $C_{jt}\to 0$ as $t\to\infty$. Moreover, $0\le C_{jt}\le C_j$ for every $j,t$. Since $\sum_{j=1}^\infty C_j<\infty$, the dominated convergence theorem applied to the counting measure on $\N$ implies
\begin{equation*}
    \lim_{t\to\infty}\sum_{j=1}^t\pi^{j-1}(1-\pi)\E_0[\widehat Q_t\mid \tau=j]
    =\lim_{t\to\infty}\sum_{j=1}^\infty C_{jt}=0.
\end{equation*}
Noting that $\widehat Q_t=m_{0\to t}Q_t$, letting $t\to\infty$ in \eqref{eq:mQ_iter}, the no-bubble condition holds if and only if \eqref{eq:TVC} holds.

Finally, if $(e_t,D_t)$ is $\cF_{t-1}$-measurable or $\gamma=1$, the stochastic discount factor \eqref{eq:SDF_Q} reduces to $m_{t\to t+1}=Q_t/(Q_{t+1}+D_{t+1})$. Therefore, as in \eqref{eq:mQ_b},
\begin{equation*}
    \widehat Q_t=Q_0\prod_{s=1}^t\frac{Q_s}{Q_s+D_s}\le Q_0.
\end{equation*}
It follows that $S_t\le \pi^tQ_0\to 0$, so \eqref{eq:TVC} holds and there is no bubble. \hfill \qedsymbol

\subsection{Proof of Proposition \ref{prop:eq_GH}}

Fix a date $t$. Given $\phi_t\in(0,1]$, the skilled-labor market clearing condition \eqref{eq:laborH} uniquely determines $X_t$. The final-good firm's first-order conditions \eqref{eq:Pt} and \eqref{eq:wLt} then determine $P_t$ and $w_{Lt}$, while \eqref{eq:ptj} gives $w_{Ht}=(\sigma-1)P_t/\sigma$. Thus \eqref{eq:eqprice_P}--\eqref{eq:eqprice_wL} hold.

Suppose first that innovation occurs, so $\phi_t<1$. Occupational indifference implies $Q_t=w_{Ht}/(an_t)$. Together with \eqref{eq:stock_demand_GH}, this gives
\begin{equation*}
    \theta_{Ht}=\beta an_t,\qquad
    \theta_{Lt}=\frac{w_{Lt}}{w_{Ht}}\beta an_t.
\end{equation*}
Substituting these expressions into \eqref{eq:n_dyn} and \eqref{eq:n_clear} gives
\begin{equation*}
    (1+a(1-\phi_t)H)n_t=H\beta an_t
    +L\frac{\sigma F_LA_{Lt}}{(\sigma-1)F_XA_{Xt}}\beta an_t,
\end{equation*}
where $F_X,F_L$ are evaluated at $(A_{Xt}X_t,A_{Lt}L)=(A_{Xt}\phi_t H,A_{Lt}L)$. Dividing both sides by $an_tH>0$ and using the homogeneity of $F$ and the definition of $g$, we obtain
\begin{equation*}
    \frac{1}{aH}+1-\phi_t=\beta
    +\frac{\beta\sigma}{(\sigma-1)\zeta_tg(\zeta_t\phi_t)},
\end{equation*}
which is equivalent to \eqref{eq:eqcond1}. Concavity of $F$ implies that $g$ is weakly decreasing. Since $g((0,\infty))=(0,\infty)$, the right-hand side of \eqref{eq:eqcond1} is strictly increasing in $\phi_t$ and converges to $\beta-1<0$ as $\phi_t\downarrow0$. Therefore, a unique solution $\phi_t\in(0,1)$ exists if and only if the right-hand side at $\phi_t=1$ strictly exceeds $1/(aH)$, which is precisely \eqref{eq:phi_cond}. In that case, the first line of \eqref{eq:eqprice_Q} follows from occupational indifference.

If \eqref{eq:phi_cond} does not hold, set $\phi_t=1$, so $n_{t+1}=n_t$ and there is no innovation. Stock-market clearing and \eqref{eq:stock_demand_GH} uniquely determine
\begin{equation*}
    Q_t=\frac{\beta(Hw_{Ht}+Lw_{Lt})}{n_t}.
\end{equation*}
At $\phi_t=1$, the no-entry condition $an_tQ_t\le w_{Ht}$ is equivalent to
\begin{equation*}
    \beta+\beta\frac{Lw_{Lt}}{Hw_{Ht}}\le\frac{1}{aH},
\end{equation*}
which, using the firms' first-order conditions and the definition of $g$, is exactly the failure of \eqref{eq:phi_cond}. Hence occupational choice is optimal in the no-innovation equilibrium as well. The condition determining the regime and, when innovation occurs, equation \eqref{eq:eqcond1}, depend on $t$ only through $\zeta_t$, so the same is true of $\phi_t$.

In both cases, the consumption rules and share demands solve the agents' problems; stock and labor markets clear by construction. The zero-profit conditions for the competitive firms, distribution of monopoly profits as dividends, agents' budget constraints, and financial-market clearing then imply final-good market clearing by Walras' law. Thus the constructed allocation is the unique rational expectations equilibrium. \hfill \qedsymbol

\subsection{Proof of Proposition \ref{prop:growth}}

In state $u$, the equilibrium condition \eqref{eq:eqcond2} reduces to
\begin{equation}
    \frac{1}{aH}=\phi_t-1+\beta
    +\frac{\beta\sigma(1-\alpha)}{(\sigma-1)\alpha}
    \left(\frac{A_XH}{A_LL}\right)^{\rho-1}
    n_t^{(\xi_u-\lambda_u)(\rho-1)}\phi_t^\rho. \label{eq:phiu_cond}
\end{equation}
At $t=0$, condition \eqref{eq:n0} implies that there exists a unique $\phi_0\in (0,1)$ solving \eqref{eq:phiu_cond}. By \eqref{eq:n_dyn}, we have $n_{t+1}>n_t$, so knowledge increases monotonically. By condition \eqref{eq:xi}, fixing $\phi_t$, the right-hand side of \eqref{eq:phiu_cond} is strictly increasing in $t$. Therefore, a unique solution $\phi_t\in (0,1)$ exists, which is strictly decreasing in $t$. Using \eqref{eq:n_dyn}, the gross growth factor of $n_t$ is bounded below by $1+a(1-\phi_0)H>1$. Therefore, $n_t\to\infty$. Letting $t\to\infty$ in \eqref{eq:phiu_cond} and using \eqref{eq:xi}, we must have $\phi_t\to 0$. Again using \eqref{eq:n_dyn}, $n_{t+1}/n_t\to G_u\coloneqq 1+aH$.

In state $b$, the innovation condition is independent of $n_t$ because $\xi_b=\lambda_b$. If \eqref{eq:phib_cond1} fails, the right-hand side of \eqref{eq:phib_cond2} is strictly increasing from $\beta-1$ as $\phi_b\downarrow0$ to a value weakly above $1/(aH)$ at $\phi_b=1$, so \eqref{eq:phib_cond2} has a unique solution $\phi_b\in(0,1]$. If \eqref{eq:phib_cond1} holds, Proposition \ref{prop:eq_GH} implies $\phi_b=1$ and no innovation occurs. Thus $\phi_t=\phi_b$ at every date in state $b$, and \eqref{eq:n_dyn} implies $n_{t+1}/n_t=G_b\coloneqq1+a(1-\phi_b)H$. \hfill \qedsymbol

\subsection{Proof of Theorem \ref{thm:bubble_GH}}\label{app:proof_bubble_GH}

For either type $i\in\set{H,L}$, the consumption and portfolio rules imply
\begin{equation*}
    c_{it}^y=(1-\beta)w_{it},\qquad
    c_{i,t+1}^o=\frac{Q_{t+1}+D_{t+1}}{Q_t}\beta w_{it}.
\end{equation*}
Substituting these expressions into the SDF \eqref{eq:SDF}, the wage $w_{it}$ cancels and we obtain the SDF \eqref{eq:SDF_Q}. Because uncertainty is driven only by $z_t\in\set{u,b}$, Assumption \ref{asmp:conditional} holds.

First consider state $b$. By Proposition \ref{prop:growth}, $\phi_t$ is constant at $\phi_b$. If innovation continues, the per-share price and dividend change by the common gross factor $G_b^{\lambda_b-1}$; if innovation stops, both are constant. Thus $D_t/Q_t$ is a positive constant. Because all uncertainty is resolved, the SDF reduces to $m_{t\to t+1}=Q_t/(Q_{t+1}+D_{t+1})$, and the same argument as in Proposition \ref{prop:nobubble_b} implies $Q_t=V_t$. The price-dividend ratio is constant by \eqref{eq:divyield2}.

It remains to consider state $u$. By \eqref{eq:n_dyn}, $n_t$ is determined at $t-1$, so it does not require the superscript $z$. Using $\phi_t\ge0$ and $G_u\coloneqq1+aH>1$, we obtain
\begin{equation}
    n_t\le n_0G_u^t. \label{eq:nt_ub}
\end{equation}
By Proposition \ref{prop:growth}\ref{item:growth_u}, $\phi_t\to0$. Therefore, for any $\epsilon>0$, there exists $n_\epsilon>0$ such that
\begin{equation}
    n_t\ge n_\epsilon(G_u-\epsilon)^t \label{eq:nt_lb}
\end{equation}
for all $t$. Rearranging \eqref{eq:eqcond2} in state $u$ gives
\begin{equation*}
    \frac{\beta\sigma(1-\alpha)}{(\sigma-1)\alpha}
    \left(\frac{A_XH}{A_LL}\right)^{\rho-1}
    n_t^\kappa(\phi_t^u)^\rho
    =\frac{1}{aH}+1-\beta-\phi_t^u.
\end{equation*}
It follows that there exist constants $C_\phi,\bar C_\phi>0$ such that
\begin{equation}
    \phi_t^u\sim C_\phi n_t^{-\kappa/\rho},\qquad
    \phi_t^u\le\bar C_\phi n_t^{-\kappa/\rho}. \label{eq:phit_ub}
\end{equation}
Choose $\epsilon>0$ such that $G_u-\epsilon>1$. Combining \eqref{eq:divyield2}, \eqref{eq:phit_ub}, and \eqref{eq:nt_lb}, we obtain
\begin{equation*}
    \frac{D_t^u}{Q_t^u}\le \frac{aH}{\sigma-1}\bar C_\phi n_t^{-\kappa/\rho}
    \le C(G_u-\epsilon)^{-(\kappa/\rho)t}.
\end{equation*}
Therefore, $\sum_{t=1}^\infty D_t^u/Q_t^u<\infty$, so the first condition in \eqref{eq:finite_sum} holds. Moreover, \eqref{eq:divyield2} and \eqref{eq:phit_ub} imply that $Q_t^u/D_t^u$ asymptotically grows at an exponential rate.

Conditional on $z_0=\dots=z_{t-1}=u$ and $z_t=b$, we have $n_t\sim G_u^t$ and $\phi_t=\phi_b$. Since $\lambda_b=\xi_b$, \eqref{eq:Yt_CES} implies
\begin{equation}
    Y_t^b\sim n_t^{\lambda_b}\sim G_u^{\lambda_bt}. \label{eq:Yb_order}
\end{equation}
In state $u$, \eqref{eq:phit_ub} gives $n_t^{\xi_u}\phi_t\sim n_t^{\xi_u-\kappa/\rho}$. Using the definition of $\kappa$, the exponents of $n_t$ inside the bracket of \eqref{eq:Yt_CES} are ordered according to
\begin{equation*}
    (1-\rho)(\xi_u-\kappa/\rho)\gtrless(1-\rho)\lambda_u
    \iff -\rho\kappa\gtrless(1-\rho)\kappa
    \iff \kappa\lessgtr0.
\end{equation*}
Since $\kappa>0$ and $n_t\sim G_u^t$, we obtain
\begin{equation}
    Y_t^u\sim n_t^{\lambda_u}\sim G_u^{\lambda_ut}. \label{eq:Yu_order}
\end{equation}
While state $u$ persists, innovation occurs. Hence the first line of \eqref{eq:eqprice_Q}, the CES production function, and \eqref{eq:laborH} imply
\begin{equation*}
    Q_t^u=\frac{(\sigma-1)\alpha}{\sigma an_t}(Y_t^u)^\rho
    (A_Xn_t^{\xi_u})^{1-\rho}(\phi_t^uH)^{-\rho}
    \sim n_t^{\lambda_u-1}.
\end{equation*}
Upon a transition to state $b$, either line of \eqref{eq:eqprice_Q} may apply. If innovation continues, the first line gives $Q_t^b\sim n_t^{\lambda_b-1}$. If innovation stops, the second line gives the same order because both wages are of order $n_t^{\lambda_b}$. Therefore, $Q_t^b/Q_t^u\sim n_t^{\lambda_b-\lambda_u}$. The same order holds with $Q_t+D_t$ in place of $Q_t$ because
\begin{equation*}
    \frac{Q_t^b+D_t^b}{Q_t^u+D_t^u}
    =\frac{Q_t^b}{Q_t^u}
    \frac{1+D_t^b/Q_t^b}{1+D_t^u/Q_t^u},
\end{equation*}
$D_t^b/Q_t^b$ is constant, and $D_t^u/Q_t^u\to0$. Since $\lambda_u>\lambda_b$, $n_t$ grows exponentially, and $\gamma<1$, the second condition in \eqref{eq:finite_sum} holds. Theorem \ref{thm:bubble_toy} therefore implies $Q_t>V_t$ in state $u$. The same argument applies after every history ending in state $u$. \hfill \qedsymbol

\subsection{Proof of Proposition \ref{prop:IPO_capture}}

Fix the benchmark values of $n_t$, $A_{Xt}$, $A_{Lt}$, $V_t$, and $B_t$. Under the CES specification, the skilled wage
\begin{equation*}
    w_{Ht}(\phi)=\frac{\sigma-1}{\sigma}A_{Xt}F_X(A_{Xt}H\phi,A_{Lt}L)
\end{equation*}
is continuous and strictly decreasing in $\phi\in(0,1]$. At $\chi=1$, \eqref{eq:PE_occupation} is the equilibrium occupational choice condition. Condition \eqref{eq:n0} implies active innovation in state $u$, so $\phi_t(1)\in(0,1)$ and
\begin{equation*}
    w_{Ht}(\phi_t(1))=an_t(V_t+B_t).
\end{equation*}
For any $\chi\in[0,1]$, if $an_t(V_t+\chi B_t)\le w_{Ht}(1)$, the unique solution to \eqref{eq:PE_occupation} is $\phi_t(\chi)=1$. Otherwise, continuity and strict monotonicity imply that there is a unique $\phi_t(\chi)\in[\phi_t(1),1)$ satisfying
\begin{equation*}
    w_{Ht}(\phi_t(\chi))=an_t(V_t+\chi B_t).
\end{equation*}
It follows that $\phi_t(\chi)$ decreases with $\chi$ because $B_t>0$ by Theorem \ref{thm:bubble_GH}.

Because $w_{Ht}(\phi)$ is strictly decreasing, the skilled wage increases as $\phi_t(\chi)$ falls. The CES production function satisfies $F_{XL}>0$, so
\begin{equation*}
    w_{Lt}(\phi)=A_{Lt}F_L(A_{Xt}H\phi,A_{Lt}L)
\end{equation*}
is strictly increasing in $\phi$ and hence decreases with $\chi$. More explicitly,
\begin{equation*}
    \frac{w_{Lt}(\phi)}{w_{Ht}(\phi)}
    =\frac{\sigma(1-\alpha)}{(\sigma-1)\alpha}
    (A_{Xt}/A_{Lt})^{\rho-1}
    (H/L)^\rho\phi^\rho,
\end{equation*}
so the skilled-to-unskilled wage ratio increases as $\phi$ falls. Since $F_X>0$, output $Y_t(\phi)=F(A_{Xt}H\phi,A_{Lt}L)$ is strictly increasing in $\phi$ and therefore decreases with $\chi$. Finally, R\&D employment is $H(1-\phi_t(\chi))$, and \eqref{eq:PE_knowledge} is decreasing in $\phi_t(\chi)$. These observations establish all the claims. \hfill \qedsymbol

\subsection{Proof of Proposition \ref{prop:transition_consequences}}

\ref{item:transition1} Equations \eqref{eq:Yb_order} and \eqref{eq:Yu_order} imply $Y_t^b\sim n_t^{\lambda_b}$ and $Y_{t-1}^u\sim n_{t-1}^{\lambda_u}$. While state $u$ persists, innovation occurs, so the first line of \eqref{eq:eqprice_Q}, the CES production function, and \eqref{eq:laborH} imply $Q_{t-1}^u\sim n_{t-1}^{\lambda_u-1}$. Upon transition to state $b$, either line of \eqref{eq:eqprice_Q} may apply. If innovation continues, the first line implies $Q_t^b\sim n_t^{\lambda_b-1}$; if innovation stops, the second line gives the same order because both wages are of order $n_t^{\lambda_b}$. Finally, Proposition \ref{prop:growth} gives $n_t/n_{t-1}\to G_u$. Therefore, $Y_t^b/Y_{t-1}^u\sim n_t^{\lambda_b-\lambda_u}$ and $Q_t^b/Q_{t-1}^u\sim n_t^{\lambda_b-\lambda_u}$. Since $\lambda_b<\lambda_u$, \eqref{eq:transition_declines} holds.

\medskip
\noindent \ref{item:transition2} Once state $b$ is reached, Proposition \ref{prop:growth} implies $n_{\tau+k}=G_b^kn_\tau$ for $k\ge 0$. Homogeneity of $F$ and the degree-zero homogeneity of its partial derivatives then give \eqref{eq:post_transition_levels}, with
\begin{align*}
    C_Y&=F(A_XH\phi_b,A_LL),\\
    C_H&=\frac{\sigma-1}{\sigma}A_XF_X(A_XH\phi_b,A_LL),\\
    C_L&=A_LF_L(A_XH\phi_b,A_LL).
\end{align*}
Knowledge $n_\tau$ strictly increases with $\tau$ because innovation is active throughout state $u$. Therefore, the three level paths increase with $\tau$, strictly if $\lambda_b>0$, whereas the wage ratio $w_{H,\tau+k}/w_{L,\tau+k}=C_H/C_L$ is independent of $\tau$ and $k$. \hfill \qedsymbol

\subsection{Proof of Proposition \ref{prop:knife-edge}}

At any relative productivity for which innovation occurs, \eqref{eq:divyield1} and \eqref{eq:laborH} imply
\begin{equation*}
    \frac{D_t}{Q_t}=\frac{aH}{\sigma-1}\phi_t.
\end{equation*}
Therefore, the price-dividend ratio is invariant to relative productivity if and only if $\phi_t$ is invariant. Let this common value be $\phi\in(0,1)$ and set $x\coloneqq\zeta_t\phi$. The equilibrium condition \eqref{eq:eqcond1} can be rearranged as
\begin{equation*}
    \frac{1}{\phi}\left(\frac{1}{aH}+1-\beta\right)
    =1+\frac{\beta\sigma}{(\sigma-1)xg(x)}.
\end{equation*}
As $A_{Xt}/A_{Lt}$ ranges over $(0,\infty)$, so does $x$. Hence invariance of $\phi$ implies that $xg(x)=c$ for every $x>0$ and some $c>0$.

Let $f(x)\coloneqq F(x,1)$. Homogeneity of $F$ implies $F_X(x,1)=f'(x)$ and $F_L(x,1)=f(x)-xf'(x)$. Consequently,
\begin{equation*}
    xg(x)=x\frac{F_X(x,1)}{F_L(x,1)}
    =\frac{xf'(x)}{f(x)-xf'(x)}=c.
\end{equation*}
Rearranging yields
\begin{equation*}
    \frac{f'(x)}{f(x)}=\frac{c}{(1+c)x}\eqqcolon\frac{\alpha}{x},
\end{equation*}
where $\alpha\in(0,1)$. It follows that $f(x)=Ax^\alpha$ for some $A>0$, and therefore
\begin{equation*}
    F(X,L)=Lf(X/L)=AX^\alpha L^{1-\alpha},
\end{equation*}
so $F$ is Cobb--Douglas. Conversely, if $F$ is Cobb--Douglas, then $xg(x)=\alpha/(1-\alpha)$ for every $x>0$. The equilibrium condition \eqref{eq:eqcond1} is then independent of relative productivity, so uniqueness of the interior solution implies that $\phi_t$, and hence $Q_t/D_t$, is invariant to relative productivity.

For the CES production function \eqref{eq:F_CES}, \eqref{eq:g_CES} gives
\begin{equation*}
    xg(x)=\frac{\alpha}{1-\alpha}x^{1-\rho}.
\end{equation*}
If $(A_{Xt},A_{Lt})=(A_Xn_t^\xi,A_Ln_t^\lambda)$, then the equilibrium condition is
\begin{equation*}
    \frac{1}{aH}=\phi_t-1+\beta
    +\frac{\beta\sigma(1-\alpha)}{(\sigma-1)\alpha}
    \left(\frac{A_XH}{A_LL}\right)^{\rho-1}
    n_t^{(\xi-\lambda)(\rho-1)}\phi_t^\rho.
\end{equation*}
Innovation implies $n_{t+1}>n_t$. If $(\xi-\lambda)(\rho-1)=0$, the unique interior solution $\phi_t$ is constant. If $(\xi-\lambda)(\rho-1)\ne0$, a constant $\phi_t$ cannot satisfy this equation at two distinct values of $n_t$. The dividend-yield formula above therefore proves the second claim. \hfill \qedsymbol

\printbibliography

\clearpage

\begin{refsection}

\begin{center}
{\Large\bfseries Online Appendix (not for publication)}
\end{center}

\section{Additional results and proofs}\label{sec:additional_results}

\subsection{Proof of Theorem \ref{thm:bubble_BY}}\label{app:proof_bubble_BY}

In what follows, we maintain the assumptions of Theorem \ref{thm:bubble_BY}.

We first derive the equilibrium conditions used in the proof. Let $W_{i,\tau,t}$ denote the wealth before consumption of an agent of type $i$ born at date $\tau$. A newborn agent has $W_{i,t,t}=w_{it}$, whereas an older agent holding $\theta_{i,\tau,t-1}$ shares has $W_{i,\tau,t}=(Q_t+D_t)\theta_{i,\tau,t-1}$. Assumption \ref{asmp:utility_BY} implies $c_{i,\tau,t}=(1-\beta)W_{i,\tau,t}$ and a value function of the form
\begin{equation}
    U_{i,\tau,t}=\mu_tW_{i,\tau,t}. \label{eq:value_BY}
\end{equation}
Define the constant $C$ and the gross stock return $R_{t+1}$ by
\begin{equation}
    C\coloneqq(1-\beta)^{1-\beta}\beta^\beta,
    \qquad
    R_{t+1}\coloneqq\frac{Q_{t+1}+D_{t+1}}{Q_t}. \label{eq:CR_BY}
\end{equation}
Optimal portfolio choice gives the continuation-value recursion and SDF
\begin{align}
    \mu_t
    &=C\E_t[(\mu_{t+1}R_{t+1})^{1-\gamma}]^{\frac{\beta}{1-\gamma}},
    \label{eq:mu_recursion_BY}\\
    m_{t\to t+1}
    &=\frac{\mu_{t+1}^{1-\gamma}R_{t+1}^{-\gamma}}
    {\E_t[(\mu_{t+1}R_{t+1})^{1-\gamma}]}.
    \label{eq:SDF_BY}
\end{align}
At date $t$, aggregate wealth is $Hw_{Ht}+Lw_{Lt}+(Q_t+D_t)n_t$. Because every agent saves the fraction $\beta$ of wealth, stock-market clearing requires
\begin{equation}
    Q_tn_{t+1}
    =\beta\left[Hw_{Ht}+Lw_{Lt}+(Q_t+D_t)n_t\right].
    \label{eq:asset_mkt_BY}
\end{equation}
Under active innovation, \eqref{eq:indifference} further implies $Q_t=w_{Ht}/(an_t)$.

We divide the proof into three lemmas: the first determines the equilibrium allocation and stock price, the second selects the continuation-value solution, and the third characterizes bubbles using the resulting SDF.

\begin{lem}[Equilibrium allocation]\label{lem:real_BY}
At every date and after every history, the equilibrium allocation and stock price are uniquely determined. The fraction of skilled agents working in intermediate-good production is the unique solution $\phi_t\in(0,1)$ to
\begin{equation}
    Kn_t^{\kappa_{z_t}}\phi_t^\rho+\eta\phi_t=M,
    \qquad
    \kappa_u\coloneqq\kappa,\quad \kappa_b\coloneqq0,
    \label{eq:macro_dynamics_BY}
\end{equation}
and knowledge evolves according to \eqref{eq:n_dyn}. Conditional on remaining in state $u$, $\phi_t$ decreases to zero,
\begin{equation}
    \phi_t\sim (M/K)^{1/\rho}n_t^{-\kappa/\rho},
    \qquad
    \frac{n_{t+1}}{n_t}\to G_u\coloneqq1+aH.
    \label{eq:phi_order_BY}
\end{equation}
In state $b$, $\phi_t=\phi_b\in(0,1)$ is constant and $n_{t+1}/n_t=G_b\coloneqq1+a(1-\phi_b)H$.
\end{lem}

\begin{proof}
Substitute $n_{t+1}/n_t=1+a(1-\phi_t)H$, $Q_t=w_{Ht}/(an_t)$, and $D_t/Q_t=aH\phi_t/(\sigma-1)$ into \eqref{eq:asset_mkt_BY}. Using the CES wage ratio then yields \eqref{eq:macro_dynamics_BY}. Its left-hand side is strictly increasing in $\phi_t$, equals zero at $\phi_t=0$, and exceeds $M$ at $\phi_t=1$ by \eqref{eq:active_BY}. Hence it has a unique solution in $(0,1)$. In state $u$, this solution is strictly decreasing in $n_t$. Because $n_{t+1}>n_t$, it follows that $\phi_t$ decreases. Knowledge grows at a rate bounded strictly above one, so $n_t\to\infty$ and \eqref{eq:macro_dynamics_BY} implies $\phi_t\to0$. Dividing that equation by $Kn_t^\kappa$ gives the first limit in \eqref{eq:phi_order_BY}; the second follows from \eqref{eq:n_dyn}. In state $b$, \eqref{eq:macro_dynamics_BY} is independent of $n_t$, which proves the remaining claims.
\end{proof}

In the absorbing state $b$, both $Q_t$ and $D_t$ change by the gross factor $G_b^{\lambda_b-1}$, so the stock return is the positive constant $R_b=G_b^{\lambda_b-1}(1+D_t^b/Q_t^b)$. Consequently, the unique positive stationary continuation-value coefficient is $\mu_b=(C R_b^\beta)^{1/(1-\beta)}$, where $C$ is as in \eqref{eq:CR_BY}.

\begin{lem}[Continuation-value saddle path]\label{lem:stable_BY}
There exists a unique admissible solution to \eqref{eq:mu_recursion_BY}. Along a history that remains in state $u$,
\begin{equation}
    \mu_t^u\to
    \mu_u\coloneqq
    \left(C\pi^{\frac{\beta}{1-\gamma}}R_u^\beta\right)^{\frac{1}{1-\beta}},
    \qquad
    R_u\coloneqq G_u^{\lambda_u-1}.
    \label{eq:muu_BY}
\end{equation}
\end{lem}

\begin{proof}
Define the normalized state vector
\begin{equation}
    v_t=(v_{1,t},v_{2,t},v_{3,t})
    \coloneqq
    \left(n_t^{-\kappa/\rho},
    n_t^{-\delta},
    (\mu_t^u)^{1-\gamma}\right),
    \label{eq:xi_BY}
\end{equation}
where $\delta\coloneqq(\lambda_u-\lambda_b)(1-\gamma)>0$, and let $v_3^b\coloneqq\mu_b^{1-\gamma}$. Conditional on state $u$ at date $t$, write $R_{t+1}^{uu}$ and $R_{t+1}^{ub}$ for the stock return when the next state is $u$ and $b$, respectively. Equation \eqref{eq:mu_recursion_BY} is equivalent to
\begin{equation}
    v_{3,t+1}
    =
    \frac{C^{-(1-\gamma)/\beta}v_{3,t}^{1/\beta}
    -(1-\pi)v_3^b(R_{t+1}^{ub})^{1-\gamma}}
    {\pi(R_{t+1}^{uu})^{1-\gamma}}.
    \label{eq:xi3_forward_BY}
\end{equation}

We put this nonautonomous recursion into an autonomous system near its limiting steady state. By \eqref{eq:macro_dynamics_BY} and the implicit function theorem, there is a positive continuously differentiable function $h$ near zero such that $\phi_t=v_{1,t}h(v_{1,t})$ and $h(0)=(M/K)^{1/\rho}$. Writing $G(v_1)\coloneqq1+aH(1-v_1h(v_1))$, the first two components of \eqref{eq:xi_BY} obey
\begin{equation}
    v_{1,t+1}=v_{1,t}G(v_{1,t})^{-\kappa/\rho},
    \qquad
    v_{2,t+1}=v_{2,t}G(v_{1,t})^{-\delta}.
    \label{eq:xi12_BY}
\end{equation}
Direct substitution into the CES price and dividend formulas shows that there exist positive continuously differentiable functions $r_u,r_b$ near zero such that
\begin{equation}
    R_{t+1}^{uu}=r_u(v_{1,t}),\qquad
    (R_{t+1}^{ub})^{1-\gamma}=v_{2,t}r_b(v_{1,t}),
    \qquad r_u(0)=R_u.
    \label{eq:returns_normalized_BY}
\end{equation}
For completeness, this follows from $Q_t^u=n_t^{\lambda_u-1}q(v_{1,t})$ for a positive continuously differentiable function $q$, $D_t^u/Q_t^u=aHv_{1,t}h(v_{1,t})/(\sigma-1)$, and $Q_{t+1}^b/Q_t^u=n_t^{\lambda_b-\lambda_u}$ times a positive continuously differentiable function of $v_{1,t}$.

Equations \eqref{eq:xi3_forward_BY}--\eqref{eq:returns_normalized_BY} define a continuously differentiable autonomous map in $v=(v_1,v_2,v_3)$ near the fixed point $v^*\coloneqq(0,0,v_3^*)$, where $v_3^*\coloneqq\mu_u^{1-\gamma}$. The eigenvalues of its Jacobian at $v^*$ are $G_u^{-\kappa/\rho}$, $G_u^{-\delta}$, and $1/\beta$. The first two lie strictly inside the unit circle, whereas $1/\beta>1$. The fixed point is therefore hyperbolic with a two-dimensional stable subspace. In economic terms, $v_1$ and $v_2$ are predetermined technological state variables, whereas $v_3$ is forward-looking. By the local stable manifold theorem \citep[Theorem~8.9]{TodaEME}, the local stable set is two-dimensional and every point on it converges to $v^*$. Because the stable subspace projects nonsingularly onto the $(v_1,v_2)$ coordinates, the local stable set can be expressed as a graph $v_3=\mathcal S(v_1,v_2)$. Thus, for each technological state $(v_1,v_2)$ sufficiently close to $(0,0)$, there is a unique local choice of $v_3$ that converges to the fixed point.

Since $n_t\to\infty$, there exists $T$ such that $(v_{1,T},v_{2,T})$ belongs to the domain of this graph. Set $v_{3,T}=\mathcal S(v_{1,T},v_{2,T})$, generate the continuation on the local stable manifold, and extend the solution uniquely backward from $T$ using the original recursion
\begin{equation*}
    v_{3,t}=C^{1-\gamma}
    \left[\pi v_{3,t+1}(R_{t+1}^{uu})^{1-\gamma}
    +(1-\pi)v_3^b(R_{t+1}^{ub})^{1-\gamma}\right]^\beta.
\end{equation*}
This constructs a positive solution satisfying \eqref{eq:admissible_BY}.

To prove uniqueness, take any admissible solution and set $y_t\coloneqq\log(v_{3,t}/v_3^*)$. The fixed point satisfies
\begin{equation*}
    v_3^*
    =\frac{C^{-(1-\gamma)/\beta}(v_3^*)^{1/\beta}}
    {\pi R_u^{1-\gamma}}.
\end{equation*}
Dividing \eqref{eq:xi3_forward_BY} by this equality gives the exact identity
\begin{equation*}
    \frac{v_{3,t+1}}{v_3^*}
    =\left(\frac{v_{3,t}}{v_3^*}\right)^{1/\beta}
    \left(\frac{R_u}{R_{t+1}^{uu}}\right)^{1-\gamma}
    \left[
    1-\frac{(1-\pi)v_3^b(R_{t+1}^{ub})^{1-\gamma}}
    {C^{-(1-\gamma)/\beta}v_{3,t}^{1/\beta}}
    \right].
\end{equation*}
Because the solution is positive, the expression in square brackets is positive. Taking logs therefore yields
\begin{equation}
    y_{t+1}=\frac{1}{\beta}y_t+\epsilon_t, \label{eq:yt_BY}
\end{equation}
where
\begin{equation*}
    \epsilon_t\coloneqq
    (1-\gamma)\log\left(\frac{R_u}{R_{t+1}^{uu}}\right)
    +\log\left[
    1-\frac{(1-\pi)v_3^b(R_{t+1}^{ub})^{1-\gamma}}
    {C^{-(1-\gamma)/\beta}v_{3,t}^{1/\beta}}
    \right].
\end{equation*}
Because $R_{t+1}^{uu}\to R_u$, $(R_{t+1}^{ub})^{1-\gamma}\to0$, and admissibility bounds $v_{3,t}$ away from zero, the sequence $y_t$ is bounded and $\epsilon_t\to0$. Iterating \eqref{eq:yt_BY}, we obtain
\begin{equation*}
    y_t=-\sum_{s=t}^\infty\beta^{s-t+1}\epsilon_s,
    \qquad
    \abs{y_t}\le
    \frac{\beta}{1-\beta}\sup_{s\ge t}\abs{\epsilon_s}\to 0.
\end{equation*}
Thus $v_{3,t}\to v_3^*$. Hence every admissible solution eventually lies on the local stable manifold. The graph property and backward recursion then imply uniqueness.
\end{proof}

The unique admissible sequence determines the common SDF in \eqref{eq:SDF_BY}. Every agent's resulting plan also satisfies the transversality condition. Indeed, optimal portfolio choice gives $\E_t[m_{t\to t+1}R_{t+1}]=1$, while next-period wealth equals $\beta W_{i,\tau,t}R_{t+1}$. Iterating conditional expectations yields
\begin{equation}
    \E_t[m_{t\to t+k}W_{i,\tau,t+k}]
    =\beta^k W_{i,\tau,t}\to 0
    \label{eq:TVC_BY}
\end{equation}
as $k\to\infty$. Along a history that has remained in state $u$ through date $t-1$, define the two possible cum-dividend stock payoffs at $t$ by $\Pi_t^z\coloneqq Q_t^z+D_t^z$ for $z\in\set{u,b}$.

\begin{lem}[Bubble characterization]\label{lem:bubble_BY}
Suppose $z_0=u$ and let $\set{\mu_t}_{t=0}^\infty$ be the unique admissible continuation-value solution. There is a bubble at date 0 if and only if
\begin{equation}
    \sum_{t=1}^\infty\frac{D_t^u}{Q_t^u}<\infty
    \quad\text{and}\quad
    \sum_{t=1}^\infty
    \left(\frac{\Pi_t^b}{\Pi_t^u}\right)^{1-\gamma}<\infty.
    \label{eq:bubble_conditions_BY}
\end{equation}
\end{lem}

\begin{proof}
State $b$ is deterministic and has a positive constant dividend yield. The same iteration as in Proposition \ref{prop:nobubble_b} therefore shows that it is bubbleless. It remains to evaluate the discounted terminal price on histories that stay in state $u$. From \eqref{eq:SDF_BY}, their contribution through date $T$ is
\begin{equation*}
    Q_0\prod_{t=1}^T
    \left[
    1+\frac{D_t^u}{Q_t^u}
    +\frac{1-\pi}{\pi}\frac{\Pi_t^u}{Q_t^u}
    \left(\frac{\mu_b\Pi_t^b}{\mu_t^u\Pi_t^u}\right)^{1-\gamma}
    \right]^{-1}.
\end{equation*}
This product has a strictly positive limit if and only if the sum of the two nonnegative terms inside the brackets is finite. If the first series in \eqref{eq:bubble_conditions_BY} converges, then $\Pi_t^u/Q_t^u=1+D_t^u/Q_t^u$ is bounded. Because $\mu_t^u\to\mu_u\in(0,\infty)$, finiteness is equivalent to the two conditions in \eqref{eq:bubble_conditions_BY}.
\end{proof}

\begin{proof}[Proof of Theorem \ref{thm:bubble_BY}]
Lemma \ref{lem:real_BY} establishes uniqueness of the equilibrium allocation and stock price, and Lemma \ref{lem:stable_BY} establishes uniqueness of the admissible continuation-value solution and its associated SDF. In state $u$,
\begin{equation*}
    \frac{D_t^u}{Q_t^u}
    =\frac{aH}{\sigma-1}\phi_t
    \sim \frac{aH}{\sigma-1}(M/K)^{1/\rho}n_t^{-\kappa/\rho}.
\end{equation*}
Since $n_{t+1}/n_t\to G_u>1$, the first series in \eqref{eq:bubble_conditions_BY} converges and $Q_t^u/D_t^u$ grows exponentially. The CES price formulas and Lemma \ref{lem:real_BY} imply $\Pi_t^b/\Pi_t^u\sim n_t^{\lambda_b-\lambda_u}$. Because $\lambda_u>\lambda_b$ and $\gamma<1$, the second series in \eqref{eq:bubble_conditions_BY} also converges. Lemma \ref{lem:bubble_BY} therefore implies $Q_t>V_t$ in state $u$. The same tail argument applies after every history that remains in state $u$. In state $b$, the dividend yield is a positive constant and the argument in the proof of Lemma \ref{lem:bubble_BY} gives $Q_t=V_t$.
\end{proof}

\begin{rem}\label{rem:infinitely_lived_agents}
Continual cohort entry is central to Theorem \ref{thm:bubble_BY}. Each existing cohort gradually runs down its stock position, while every new cohort invests part of its labor income in the stock market. Entrants therefore replenish aggregate stock demand even though every agent satisfies the transversality condition \eqref{eq:TVC_BY}. The bubble relies neither on an agent maintaining an ever-growing position nor on an exogenous borrowing or short-sale constraint. This result does not contradict the deterministic no-bubble theorem of \citet{Kamihigashi2018}, which requires some agent to be able, from some date onward, to reduce asset holdings permanently by a uniformly positive amount. In our model, the end-of-period stock holdings of cohort $(i,\tau)$ satisfy $\theta_{i,\tau,t}/\theta_{i,\tau,t-1}=\beta(1+D_t/Q_t)$. Conditional on remaining in state $u$, $D_t/Q_t\to0$, so this ratio converges to $\beta<1$ and $\theta_{i,\tau,t}\to0$. Hence no cohort can reduce its position by a fixed positive amount at all sufficiently late dates while maintaining nonnegative wealth. Although the theorem of \citet{Kamihigashi2018} does not directly apply in our stochastic setting, the cohort-entry channel here is consistent with its portfolio-reduction condition.
\end{rem}

\subsection{Epstein-Zin preferences with \texorpdfstring{$\text{EIS} \neq 1$}{}}\label{app:non-unit_EIS}

In \S\ref{sec:toy}, we assumed a unit elasticity of intertemporal substitution (EIS) for analytical tractability. We now allow the EIS to differ from 1 and clarify which parts of the baseline characterization remain valid. The distinction is important because, away from unit EIS, the equilibrium saving rate and asset price are endogenous.

For any positive random variable $X$, define its conditional CRRA certainty equivalent by
\begin{equation*}
    \cR_t(X)\coloneqq
    \begin{cases*}
        \E_t[X^{1-\gamma}]^\frac{1}{1-\gamma} & if $\gamma\neq1$,\\
        \exp(\E_t[\log X]) & if $\gamma=1$.
    \end{cases*}
\end{equation*}

\begin{asmp}\label{asmp:utility_non-unit_EIS}
The utility function of an agent born at date $t$ is
\begin{equation}
    U(c_t^y,c_{t+1}^o)=\left((1-\beta)(c_t^y)^{1-1/\psi}
    +\beta\cR_t(c_{t+1}^o)^{1-1/\psi}\right)^\frac{1}{1-1/\psi},
    \label{eq:utility_non-unit_EIS}
\end{equation}
where $\beta\in(0,1)$ and $0<\psi\neq 1$ is the EIS.
\end{asmp}

For later use, we define the normalized price, gross endowment growth rate, and normalized dividend by
\begin{equation}
    x_t\coloneqq\frac{Q_t}{e_t},\qquad
    G_{t+1}\coloneqq\frac{e_{t+1}}{e_t},\qquad
    d_t\coloneqq\frac{D_t}{e_t}. \label{eq:normalized_variables_non-unit_EIS}
\end{equation}
We say that an equilibrium is \emph{uniformly interior} if there exist $0<\ubar{x}\le \bar{x}<1$ such that $x_t\in [\ubar{x},\bar{x}]$ at every history.

\begin{thm}[Existence and bubble characterization under non-unit EIS]\label{thm:bubble_interior_non-unit_EIS}
Suppose Assumptions \ref{asmp:states}--\ref{asmp:conditional} and \ref{asmp:utility_non-unit_EIS} hold. Suppose there are constants $0<\ubar{G}\le\bar{G}<\infty$ and $\bar d<\infty$ such that, at every history,
\begin{equation}
    \ubar{G}\le\cR_t(G_{t+1})\le\bar{G},
    \qquad 0<d_t\le\bar d. \label{eq:bounds_interior_existence}
\end{equation}
Then a uniformly interior rational expectations equilibrium exists. In particular:
\begin{enumerate}
    \item If $\psi<1$, every equilibrium is uniformly interior.
    \item If $\psi>1$, the set of equilibrium normalized price processes has a least and a greatest element under the pointwise order. The greatest equilibrium is the unique uniformly interior equilibrium. Every other equilibrium, if one exists, has normalized prices arbitrarily close to zero across dates and histories.
\end{enumerate}
For $z\in\set{u,b}$, let $(e_t^z,D_t^z)$ be the value of $(e_t,D_t)$ conditional on $z_0=\dots=z_{t-1}=u$ and $z_t=z$. If $z_0=u$, any uniformly interior equilibrium contains a bubble at date 0 if and only if
\begin{equation}
    \sum_{t=1}^\infty\frac{D_t^u}{e_t^u}<\infty
    \quad\text{and}\quad
    \sum_{t=1}^\infty\left(\frac{e_t^b}{e_t^u}\right)^{1-\gamma}<\infty. \label{eq:finite_sum_primitives_non-unit_EIS}
\end{equation}
\end{thm}

Note that unlike the exact consumption-based characterization in \eqref{eq:finite_sum}, condition \eqref{eq:finite_sum_primitives_non-unit_EIS} is stated entirely in terms of primitives; under \eqref{eq:bounds_interior_existence} and uniform interiority, the two formulations are equivalent. Thus, under condition \eqref{eq:finite_sum_primitives_non-unit_EIS}, Theorem \ref{thm:bubble_interior_non-unit_EIS} establishes bubble necessity despite anticipated collapse.

We prove Theorem \ref{thm:bubble_interior_non-unit_EIS} through a series of lemmas.

\begin{lem}[Equilibrium price equation and SDF]\label{lem:non-unit_EIS_equilibrium}
Suppose Assumption \ref{asmp:utility_non-unit_EIS} holds. In any rational expectations equilibrium, let
\begin{equation*}
    R_{t+1}\coloneqq\frac{Q_{t+1}+D_{t+1}}{Q_t}
    \quad\text{and}\quad
    \rho_t\coloneqq\cR_t(R_{t+1}).
\end{equation*}
Then the equilibrium price equation and SDF are
\begin{align}
    Q_t&=\frac{\beta}{1-\beta}(e_t-Q_t)^{1/\psi}
    \cR_t(Q_{t+1}+D_{t+1})^{1-1/\psi}, \label{eq:price_non-unit_EIS} \\
    m_{t\to t+1}
    &=\frac{\beta}{1-\beta}
    \left(\frac{c_{t+1}^o}{c_t^y}\right)^{-1/\psi}
    \left(\frac{c_{t+1}^o}{\cR_t(c_{t+1}^o)}\right)^{1/\psi-\gamma}=\frac{Q_t(Q_{t+1}+D_{t+1})^{-\gamma}}
    {\E_t[(Q_{t+1}+D_{t+1})^{1-\gamma}]}.
    \label{eq:SDF_Q_non-unit_EIS}
\end{align}
Conversely, any adapted process $Q$ satisfying $0<Q_t<e_t$ and \eqref{eq:price_non-unit_EIS} is an equilibrium price process.
\end{lem}

\begin{proof}
Fix the return process and let $s\in(0,e_t)$ denote saving by the young. Homogeneity of $\cR_t$ makes the saving problem equivalent to
\begin{equation*}
    \max_{s\in(0,e_t)}\frac{1}{1-1/\psi}\left((1-\beta)(e_t-s)^{1-1/\psi}
    +\beta\rho_t^{1-1/\psi}s^{1-1/\psi}\right).
\end{equation*}
The objective is strictly concave. Taking the first-order condition and rearranging yields
\begin{equation*}
	\frac{Q_t}{e_t-Q_t}=\left(\frac{\beta}{1-\beta}\right)^\psi \rho_t^{\psi-1}.
\end{equation*}
Substituting $R_{t+1}=(Q_{t+1}+D_{t+1})/Q_t$ and using the homogeneity of $\cR_t$ yields \eqref{eq:price_non-unit_EIS}. Computing the marginal rate of substitution from \eqref{eq:utility_non-unit_EIS} and using the equilibrium conditions $c_t^y=e_t-Q_t$, $c_{t+1}^o=Q_{t+1}+D_{t+1}$, and \eqref{eq:price_non-unit_EIS} gives \eqref{eq:SDF_Q_non-unit_EIS}.

For the converse, set $c_t^y=e_t-Q_t$ and $c_{t+1}^o=Q_{t+1}+D_{t+1}$ and price financial claims with the strictly positive SDF in \eqref{eq:SDF_Q_non-unit_EIS}. Equation \eqref{eq:price_non-unit_EIS} implies the consumption first-order conditions, while
\begin{equation*}
    Q_t=\E_t[m_{t\to t+1}(Q_{t+1}+D_{t+1})].
\end{equation*}
Thus the proposed consumption plan exhausts the intertemporal budget. Concavity implies optimality, and financial markets clear.
\end{proof}

When $\psi=1$, \eqref{eq:price_non-unit_EIS} reduces to $Q_t=\beta e_t$. For $\psi\neq1$, it is a forward-looking equilibrium equation and neither yields a closed-form price process nor establishes uniqueness. The absence of $\psi$ from the last expression in \eqref{eq:SDF_Q_non-unit_EIS} does not change this conclusion: once the SDF is evaluated at an equilibrium price, the Euler equation holds identically, whereas the price itself must still solve \eqref{eq:price_non-unit_EIS}.

To establish equilibrium existence, for an adapted process $x$, let
\begin{equation}
    Z_t(x)\coloneqq\cR_t\left(G_{t+1}(x_{t+1}+d_{t+1})\right). \label{eq:Ztx}
\end{equation}
Let $A\coloneqq\beta/(1-\beta)$ and $F_\psi(x)\coloneqq\frac{x^\psi}{1-x}$. By homogeneity of $\cR_t$, the equilibrium condition \eqref{eq:price_non-unit_EIS} is equivalent to
\begin{equation}
    F_\psi(x_t)=A^\psi Z_t(x)^{\psi-1}. \label{eq:normalized_price_non-unit_EIS}
\end{equation}
Because $F_\psi:(0,1)\to (0,\infty)$ is strictly increasing and onto, its inverse is well defined. Define
\begin{equation*}
    H_\psi(z)\coloneqq F_\psi^{-1}\left(A^\psi z^{\psi-1}\right),
    \qquad (Tx)_t\coloneqq H_\psi(Z_t(x)).
\end{equation*}
By \eqref{eq:normalized_price_non-unit_EIS}, the operator $T$ is increasing when $\psi>1$ and decreasing when $\psi\in (0,1)$.

\begin{lem}[Existence and equilibrium structure]\label{lem:interior_eq_non-unit_EIS}
Suppose Assumptions \ref{asmp:states}--\ref{asmp:conditional} and \ref{asmp:utility_non-unit_EIS} hold, and suppose the bounds in \eqref{eq:bounds_interior_existence} hold at every history.
Then a uniformly interior equilibrium exists. In particular:
\begin{enumerate}
    \item\label{item:interior1} If $\psi<1$, every equilibrium is uniformly interior.
    \item\label{item:interior2} If $\psi>1$, the set of equilibrium normalized price processes has a least and a greatest element under the pointwise order. The greatest equilibrium is the unique uniformly interior equilibrium, and every other equilibrium, if one exists, has normalized prices with infimum zero across dates and histories.
\end{enumerate}
\end{lem}

\begin{proof}
For each $t$, let $\cH_t$ be the finite set of regime histories $h_t=(z_0,\dots,z_t)$ that can occur, and let $\cH\coloneqq\bigcup_{t=0}^\infty\cH_t$. Thus $\cH$ is countable, and an adapted process $x$ can be identified with a collection $\set{x(h):h\in\cH}$. For $\epsilon\in(0,1/2)$, consider the complete lattice $\cX_\epsilon\coloneqq[\epsilon,1-\epsilon]^\cH$. For every $x\in\cX_\epsilon$, monotonicity and homogeneity of $\cR_t$ and \eqref{eq:Ztx} imply
\begin{equation*}
    \epsilon\ubar{G}\le Z_t(x)\le(1+\bar d)\bar{G}.
\end{equation*}
Moreover, $F_\psi(\epsilon)$ is of order $\epsilon^\psi$, whereas $F_\psi(1-\epsilon)$ is of order $\epsilon^{-1}$. If $\psi>1$, the smallest and largest possible values of $A^\psi Z_t(x)^{\psi-1}$ are of orders $\epsilon^{\psi-1}$ and $1$, respectively. If $\psi<1$, they are of orders $1$ and $\epsilon^{\psi-1}$, respectively. Consequently, for all sufficiently small $\epsilon>0$, we have
\begin{equation*}
    F_\psi(\epsilon)\le A^\psi Z_t(x)^{\psi-1}
    \le F_\psi(1-\epsilon)
\end{equation*}
uniformly over $t$, histories, and $x\in\cX_\epsilon$. Hence $T\cX_\epsilon\subseteq\cX_\epsilon$.

Suppose first that $\psi>1$. The certainty-equivalent operator $\cR_t$ is increasing and concave. The function $H_\psi$ is also increasing and strictly concave. To see the latter claim, set $p\coloneqq\psi/(\psi-1)$ and $q\coloneqq1/(\psi-1)$, so $p=q+1$. The inverse of $H_\psi$ is $H_\psi^{-1}(x)=A^{-p}x^p(1-x)^{-q}$, whose second derivative is $A^{-p}pq\,x^{q-1}(1-x)^{-p-1}>0$. Thus $H_\psi^{-1}$ is strictly convex and $H_\psi$ is strictly concave. It follows that $T$, as the composition of an increasing concave function and a concave operator, is increasing and concave on $\cX_\epsilon$.

Let $\boldsymbol 1$ denote the constant process equal to one. By monotonicity and homogeneity of $\cR_t$,
\begin{equation*}
    T(\epsilon\boldsymbol 1)
    \ge H_\psi(\epsilon\ubar G)\boldsymbol 1.
\end{equation*}
Because $H_\psi(z)\sim Az^{(\psi-1)/\psi}$ as $z\downarrow0$, we have $H_\psi(\epsilon\ubar G)/\epsilon\to\infty$ as $\epsilon\downarrow0$. Hence, after reducing $\epsilon$ if necessary, there exists $\omega\in(0,1)$ such that
\begin{equation*}
    T(\epsilon\boldsymbol 1)
    \ge \epsilon\boldsymbol 1
    +\omega\left((1-\epsilon)\boldsymbol 1-\epsilon\boldsymbol 1\right).
\end{equation*}
We may regard $\cX_\epsilon$ as an order interval in the ordered Banach space $\ell^\infty(\cH)$. Du's fixed point theorem \citep[Theorem 2.1.2]{Zhang2013} therefore implies that $T$ has a unique fixed point $x\in\cX_\epsilon$.

When $\psi<1$, endow $\cX_\epsilon$ with the product topology. This set is compact and convex, and $T$ is continuous because each coordinate depends continuously on the finitely many successor coordinates. The Schauder--Tychonoff fixed point theorem \citep[Corollary 17.56]{AliprantisBorder2006} therefore yields a fixed point $x\in\cX_\epsilon$. In either case, Lemma \ref{lem:non-unit_EIS_equilibrium} implies that the asset price process $(Q_t)$ defined by $Q_t\coloneqq x_te_t$ is an equilibrium. Taking $\ubar{x}=\epsilon$ and $\bar x=1-\epsilon$ proves the existence of a uniformly interior equilibrium.

Suppose $\psi<1$ and consider any equilibrium. Set $\bar Z\coloneqq(1+\bar d)\bar G$. Because $Z_t(x)\le\bar Z$ and $H_\psi$ is decreasing, $x_t\ge H_\psi(\bar Z)\eqqcolon\ubar{x}>0$. It follows that $Z_t(x)\ge\ubar{x}\ubar G$. Applying the decreasing function $H_\psi$ once more gives $x_t\le H_\psi(\ubar{x}\ubar G)\eqqcolon\bar x<1$. Thus every equilibrium is uniformly interior, proving part \ref{item:interior1}.

It remains to prove part \ref{item:interior2}. Suppose $\psi>1$. Extend $H_\psi$ continuously to zero by setting $H_\psi(0)=0$, let $\bar Z\coloneqq(1+\bar d)\bar G$, $\bar x\coloneqq H_\psi(\bar Z)<1$, and consider the complete lattice $\cX\coloneqq[0,\bar x]^\cH$. If $x\in\cX$, then $0\le Z_t(x)\le\bar Z$ and hence $0\le (Tx)_t\le\bar x$. Thus $T\cX\subseteq\cX$. Because $T$ is increasing, Tarski's fixed point theorem \citep[Theorem 1.11]{AliprantisBorder2006} yields a least fixed point $x^-$ and a greatest fixed point $x^+$ in $\cX$. Conversely, every equilibrium fixed point $x$ satisfies
\begin{equation*}
    x_t=H_\psi(Z_t(x))\le H_\psi(\bar Z)=\bar x,
\end{equation*}
so $x\in \cX$. Moreover, $d_t>0$ at every history implies $(T0)_t>0$, and therefore every fixed point in $\cX$ is pointwise strictly positive. Since it is also bounded above by $\bar x<1$, Lemma \ref{lem:non-unit_EIS_equilibrium} shows that every such fixed point defines an equilibrium. Hence $x^-$ and $x^+$ are respectively the least and greatest equilibrium normalized price processes. Finally, $x\in\cX$ and $x^+\ge x\ge\epsilon$, so the greatest equilibrium is uniformly interior.

It remains to prove uniqueness among uniformly interior equilibria. If $x'$ and $x''$ are two such equilibria, then, for all sufficiently small $\epsilon>0$, both belong to $\cX_\epsilon$, which is invariant under $T$, and the boundary condition used above holds. Du's fixed point theorem therefore implies $x'=x''$. Since $x^+$ is uniformly interior, it is the unique uniformly interior equilibrium. Every equilibrium is bounded above by $\bar x<1$, so any other equilibrium must fail uniform interiority only at the lower boundary; equivalently, its normalized prices have infimum zero across dates and histories.
\end{proof}

\begin{rem}\label{rem:noninterior_eq_non-unit_EIS}
Suppose $\psi>1$. Set $\bar Z\coloneqq(1+\bar d)\bar G$. Every equilibrium satisfies $Z_t(x)\le\bar Z$, so the monotonicity of $H_\psi$ implies $x_t\le H_\psi(\bar Z)<1$. If, in addition, there exists a constant $\delta>0$ such that, at every history,
\begin{equation}
    \cR_t(G_{t+1}d_{t+1})\ge\delta, \label{eq:dividend_floor_non-unit_EIS}
\end{equation}
then the monotonicity of $\cR_t$ gives $Z_t(x)\ge\delta$, and hence $x_t\ge H_\psi(\delta)>0$. Thus every equilibrium is uniformly interior under \eqref{eq:dividend_floor_non-unit_EIS}, so we obtain equilibrium uniqueness by Lemma \ref{lem:interior_eq_non-unit_EIS}\ref{item:interior2}.

Since $G_{t+1}d_{t+1}=D_{t+1}/e_t$, condition \eqref{eq:dividend_floor_non-unit_EIS} requires the conditional certainty equivalent of next period's dividend, normalized by current endowment, to be uniformly bounded away from zero. Without such a condition, an equilibrium may satisfy $\inf_t x_t=0$ because $H_\psi(z)\downarrow0$ as its risk-adjusted normalized payoff $z$ approaches zero. This possibility is consistent even with the growth bounds in \eqref{eq:bounds_interior_existence}. For example, take $\gamma<1$ and choose a sufficiently small $x_0^u$. Recursively define
\begin{equation*}
    y_t\coloneqq\left(\frac{F_\psi(x_t^u)}{A^\psi}\right)^{1/(\psi-1)},
    \qquad x_{t+1}^u=d_{t+1}^u\coloneqq\frac{y_t}{2}.
\end{equation*}
Along the continuing-$u$ branch, set $G_{t+1}^u=1$. Upon a switch to $b$, set $G_{t+1}^b=y_t$ and $x_{t+1}^b=d_{t+1}^b=1/2$, and thereafter choose a constant growth rate so that the latter normalized allocation satisfies \eqref{eq:normalized_price_non-unit_EIS}. Both successor payoffs at a $u$ node then equal $y_t$, so \eqref{eq:normalized_price_non-unit_EIS} holds, while $x_t^u\downarrow0$ for small enough $x_0^u$. Since $\pi>0$ and $\gamma<1$, $\cR_t(G_{t+1})$ remains bounded away from zero despite $G_{t+1}^b\downarrow0$. Thus the invariant-lattice argument in Lemma \ref{lem:interior_eq_non-unit_EIS} selects the unique uniformly interior fixed point but does not exclude fixed points outside that lattice that are not uniformly interior.
\end{rem}

The final lemma provides the equilibrium-by-equilibrium bubble characterization without imposing unit EIS.

\begin{lem}[Equilibrium-by-equilibrium bubble characterization]\label{lem:bubble_non-unit_EIS}
Suppose Assumptions \ref{asmp:states}--\ref{asmp:conditional} and \ref{asmp:utility_non-unit_EIS} hold. Consider a rational expectations equilibrium. For $z\in\set{u,b}$, let $Q_t^z$ be the value of $Q_t$ conditional on $z_0=\dots=z_{t-1}=u$ and $z_t=z$, and let $\Pi_t^z\coloneqq Q_t^z+D_t^z$. If $z_0=u$, then the asset price contains a bubble at date 0 if and only if
\begin{equation}
    \sum_{t=1}^\infty\frac{D_t^u}{Q_t^u}<\infty
    \quad\text{and}\quad
    \sum_{t=1}^\infty\left(\frac{\Pi_t^b}{\Pi_t^u}\right)^{1-\gamma}<\infty. \label{eq:finite_sum_non-unit_EIS}
\end{equation}
\end{lem}

\begin{proof}
First, once state $b$ is reached, the asset price equals its fundamental value even when $\psi\neq1$. To see this, suppose that state $b$ is reached at date $\tau$. Assumption \ref{asmp:balanced} implies that $D_t/e_t=D_\tau/e_\tau>0$ for all $t\ge\tau$. Positive young consumption and asset market clearing imply $0<Q_t<e_t$, and hence
\begin{equation*}
    \frac{D_t}{Q_t}>\frac{D_t}{e_t}=\frac{D_\tau}{e_\tau}>0.
\end{equation*}
Because all uncertainty is resolved after $\tau$, \eqref{eq:SDF_Q_non-unit_EIS} reduces to $m_{t\to t+1}=Q_t/(Q_{t+1}+D_{t+1})$. Therefore,
\begin{equation*}
    m_{\tau\to T}Q_T
    =Q_\tau\prod_{t=\tau+1}^T\frac{Q_t}{Q_t+D_t}
    \le Q_\tau\left(1+\frac{D_\tau}{e_\tau}\right)^{-(T-\tau)}\to0.
\end{equation*}

It follows, by the same decomposition as in the proof of Proposition \ref{prop:nobubble_toy}, that a bubble exists at date 0 if and only if $S_t\coloneqq\pi^t\E_0[m_{0\to t}Q_t\mid\tau>t]$ has a strictly positive limit. Using \eqref{eq:SDF_Q_non-unit_EIS} and the state-contingent notation in the lemma, we obtain
\begin{equation*}
    S_t=Q_0\prod_{s=1}^t
    \frac{1}{1+D_s^u/Q_s^u+\frac{1-\pi}{\pi}(\Pi_s^u/Q_s^u)(\Pi_s^b/\Pi_s^u)^{1-\gamma}}.
\end{equation*}
The rest of the proof is the same as that of Theorem \ref{thm:bubble_toy}.
\end{proof}

\begin{proof}[Proof of Theorem \ref{thm:bubble_interior_non-unit_EIS}]
Lemma \ref{lem:interior_eq_non-unit_EIS} proves the claims about equilibrium existence and structure.

Fix any uniformly interior equilibrium. Then $D_t^u/Q_t^u$ and $D_t^u/e_t^u$ differ by multiplicative factors bounded above and away from zero. Moreover,
\begin{equation*}
    \frac{\Pi_t^b}{\Pi_t^u}
    =\frac{e_t^b}{e_t^u}
    \frac{x_t^b+d_t^b}{x_t^u+d_t^u},
\end{equation*}
and the last factor is bounded above and away from zero by uniform interiority and \eqref{eq:bounds_interior_existence}. Therefore, Lemma \ref{lem:bubble_non-unit_EIS} implies that the two conditions in \eqref{eq:finite_sum_non-unit_EIS} are equivalent to those in \eqref{eq:finite_sum_primitives_non-unit_EIS}. Combining this equivalence with Lemma \ref{lem:interior_eq_non-unit_EIS} proves all the claims.
\end{proof}

\begin{rem}
Lemma \ref{lem:bubble_non-unit_EIS} is an equilibrium-by-equilibrium characterization and does not by itself establish equilibrium uniqueness or independence of prices from the EIS. Without uniform interiority, the EIS may affect the convergence of the two series in \eqref{eq:finite_sum_non-unit_EIS} through its effect on $Q_t$. Theorem \ref{thm:bubble_interior_non-unit_EIS} shows that, under the stated bounds, the primitive characterization \eqref{eq:finite_sum_primitives_non-unit_EIS} applies to every uniformly interior equilibrium. It therefore applies to every equilibrium when $\psi<1$ and to the greatest equilibrium---the unique uniformly interior equilibrium---when $\psi>1$. Other equilibria with $\psi>1$ may exist, but their normalized prices must have infimum zero.
\end{rem}

\begin{rem}
Under \eqref{eq:bounds_interior_existence}, condition \eqref{eq:dividend_floor_non-unit_EIS} is incompatible with the bubble condition \eqref{eq:finite_sum_primitives_non-unit_EIS}. To see why, consider a history on which state $u$ has continued through date $t$, and write $G_{t+1}^z=e_{t+1}^z/e_t^u$ for $z\in\set{u,b}$. The first series in \eqref{eq:finite_sum_primitives_non-unit_EIS} implies $d_t^u\to0$, while the second implies $(G_{t+1}^b/G_{t+1}^u)^{1-\gamma}=(e_{t+1}^b/e_{t+1}^u)^{1-\gamma}\to0$. If $\gamma<1$, the upper bound on $\cR_t(G_{t+1})$ bounds $G_{t+1}^u$ from above, so both $G_{t+1}^u d_{t+1}^u$ and $G_{t+1}^b d_{t+1}^b$ converge to zero. Hence $\cR_t(G_{t+1}d_{t+1})\to0$, contradicting \eqref{eq:dividend_floor_non-unit_EIS}. If $\gamma=1$, the second series in \eqref{eq:finite_sum_primitives_non-unit_EIS} is a sum of ones and cannot converge. Finally, if $\gamma>1$, condition \eqref{eq:dividend_floor_non-unit_EIS} bounds $G_{t+1}^u d_{t+1}^u$ away from zero. Since $d_{t+1}^u\to0$, we must have $G_{t+1}^u\to\infty$; the second series then implies $G_{t+1}^b/G_{t+1}^u\to\infty$, so $\cR_t(G_{t+1})\to\infty$, contradicting its upper bound in \eqref{eq:bounds_interior_existence}. Thus, when $\psi>1$, imposing \eqref{eq:dividend_floor_non-unit_EIS} to make every equilibrium uniformly interior would rule out the primitive bubble condition. This is why Theorem \ref{thm:bubble_interior_non-unit_EIS} instead focuses on the greatest equilibrium---the unique uniformly interior equilibrium---whose uniform interiority does not require \eqref{eq:dividend_floor_non-unit_EIS}.
\end{rem}

\printbibliography

\end{refsection}

\end{document}